\documentclass[aps,prb,twocolumn,superscriptaddress, showpacs]{revtex4-2}

\usepackage{graphicx}
\usepackage{dcolumn}
\usepackage{bm}
\usepackage{hyperref}
\usepackage{bm, amsmath}
\usepackage{txfonts}

\begin{document}

\title{Cholesteric Fingers from a Magnetic Perspective: Topology, Energetics, and Interactions}

\author{Takayuki Shigenaga}
\affiliation{Department of Chemistry, Faculty of Science, Hiroshima University Kagamiyama, Higashi Hiroshima, Hiroshima 739-8526, Japan}
\affiliation{International Institute for Sustainability with Knotted Chiral Meta Matter, Kagamiyama, Higashi Hiroshima, Hiroshima 739-8526, Japan} 

\author{Andrey O. Leonov}
\thanks{Corresponding author: leonov@hiroshima-u.ac.jp}
\affiliation{Department of Chemistry, Faculty of Science, Hiroshima University Kagamiyama, Higashi Hiroshima, Hiroshima 739-8526, Japan}
\affiliation{International Institute for Sustainability with Knotted Chiral Meta Matter, Kagamiyama, Higashi Hiroshima, Hiroshima 739-8526, Japan}

\date{\today}

\begin{abstract}

Chiral liquid crystals and chiral magnets host a wide variety of topological solitons governed by closely related continuum theories, namely the Frank--Oseen model for liquid crystals and the Dzyaloshinskii model for chiral magnets.
Here we exploit this correspondence to develop a unified theoretical description of cholesteric fingers in confined liquid crystals and their magnetic counterparts.
Within a continuum framework including bulk and surface anisotropies, we analyze the topology, internal structure, interactions, and collective states of the two principal finger varieties, CF--1 and CF--2.

We show that cholesteric fingers can be interpreted as composite chiral solitons built from meronic constituents.
In particular, CF--2 corresponds to a bimeron with unit topological charge, whereas CF--1 represents a topologically trivial composite object formed by two merons with identical vorticities.
From a homotopic viewpoint these textures correspond to skyrmions and droplets, respectively.
Strong homeotropic surface anchoring induces pronounced confinement effects that reshape the meron structure and redistribute the topological charge across the film thickness.

Despite their composite nature, isolated fingers embedded in the homogeneous state interact repulsively and behave as particle-like objects.
Periodic finger phases emerge when the eigen-energy of an isolated finger becomes negative, giving rise to nucleation-type phase transitions with a diverging lattice period.
The coexistence of several energetically degenerate finger varieties allows the formation of mixed periodic sequences whose number grows combinatorially and can be classified analogously to stacking polytypes in close-packed crystals.
In contrast, when embedded in the conical background the interaction between fingers becomes attractive due to the overlap of distortion regions.

Finally, we demonstrate that the film thickness strongly controls the stability and internal structure of cholesteric fingers.
At small thickness the solitons collapse at a critical point where their internal topological structure disappears, whereas at large thickness isolated bimerons exhibit bistability between surface-stabilized and bulk-like configurations.
These results establish cholesteric fingers as experimentally accessible realizations of composite chiral solitons and highlight the deep correspondence between topological textures in chiral liquid crystals and chiral magnets.

\end{abstract}

\maketitle

\section{Introduction \label{sect:intro}}

Topological solitons are spatially localized, particle-like configurations of continuous fields that arise in a wide range of physical systems, from nuclear matter to condensed-matter media.
Their stability is rooted in topological invariants, which generate energetic barriers separating these nontrivial textures from homogeneous states~\cite{manton_sutcliffe,shnir,Volovik,solitons}.
Although solitons are often introduced as isolated excitations, they generally interact
with one another, exhibiting either attractive or repulsive forces.
Through these interactions, solitons may self-organize into extended mesoscopic structures
—such as soliton clusters or soliton crystals—whose characteristic length scales far
exceed the size of individual soliton cores.
We refer to such self-organized ensembles as \emph{solitonic meta-matter}, emphasizing that their effective constituents are not atoms but topologically protected field textures that behave as emergent particles~\cite{leonov2026skyrmionium}.
The formation of solitonic meta-matter highlights the cooperative role of topology, symmetry, and interactions, and provides a platform for studying collective phenomena and responses that do not exist at the level of single solitons.

\begin{figure*}
  \centering
  \includegraphics[width=0.99\linewidth]{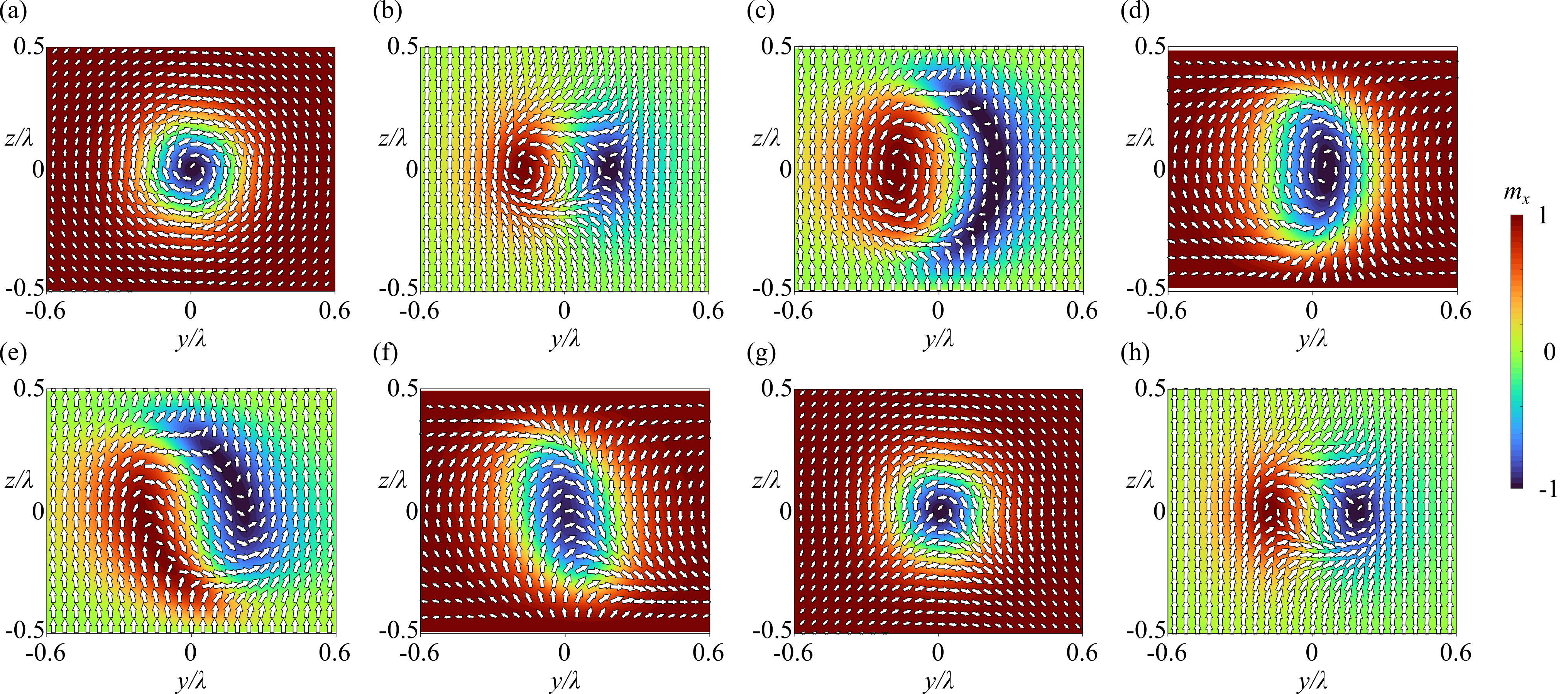}
  \caption{\label{fig01}
Axisymmetric ``relatives'' of cholesteric fingers and their topological connections.
(a) Axisymmetric chiral magnetic skyrmion embedded in the $yz$ plane, with the homogeneous
background magnetization oriented along the $x$ axis.
(b) Continuous deformation of the skyrmion into a bimeron upon reorientation of the
surrounding homogeneous state perpendicular to the skyrmion axis; the texture consists
of two coupled merons with fractional topological charges.
(c) Cholesteric finger of the second type, whose bimeron-like structure is enforced
by strong homeotropic surface anchoring; the two constituent merons possess opposite
vorticities, resulting in a distorted internal structure.
(d) CF--2 after a global rotation of the vector field by $90^\circ$ about the $y$ axis,
recovering an elliptically deformed axisymmetric skyrmion.
(e) Cholesteric finger of the first type.
(f) Axisymmetric counterpart of CF--1 obtained by rotation, revealing a composite soliton
with a rounded skyrmion-like tip and a pointed anti-skyrmion-like termination.
(g) Topologically trivial droplet formed by reversing the vorticity in one half of an
ordinary skyrmion.
(h) Reconstruction of the CF--1 texture from the droplet via an additional spin rotation.
All transformations are introduced to elucidate topological equivalence and are not meant
to imply specific physical pathways.
}
\end{figure*}

Among the many realizations of topological solitons and their associated meta-matter, two-dimensional (2D) skyrmions~\cite{JETP89,Bogdanov94} have emerged as paradigmatic examples in condensed-matter systems, including chiral magnets (ChM)~\cite{NT} and chiral liquid crystals (CLC)~\cite{Oswald}. In the typical skyrmion profile, the spins at the core
are oriented antiparallel to the surrounding uniform background and gradually
rotate to match the far-field magnetization at the outer boundary
[Fig.~\ref{fig01}(a)].

From a topological viewpoint, skyrmionic textures are maps
\(
\mathbf{m} : \mathbb{R}^2 \cup \{\infty\} \simeq S^2 \rightarrow S^2
\),
classified by the second homotopy group~\cite{Faddeev,Bott},
\(
\pi_2(S^2) \cong \mathbb{Z}
\).
This homotopy class is labeled by an integer-valued degree (winding number), which can be interpreted as the total oriented area swept out on the order-parameter sphere by the field configuration,
\begin{equation}
    Q = \frac{1}{4\pi} \int_{\mathbb{R}^2} \mathbf{m}^{\ast}(\omega_{S^2})
= \frac{1}{4\pi} \int \mathbf{m} \cdot 
\left( \partial_x \mathbf{m} \times \partial_y \mathbf{m} \right)\,
\mathrm{d}x \wedge \mathrm{d}y ,
\label{Topo}
\end{equation}
where \(\omega_{S^2}\) denotes the normalized area element  on $S^2$.
 \(\mathbf{m}\) is a smooth unit vector field, interpreted physically as the magnetization in ChM or the director field in CLC.
Skyrmionic configuration has $|Q|=1$ and corresponds to a magnetization texture that
wraps the entire unit sphere precisely once, thereby providing a clear geometric
interpretation of its topological character~\cite{NT,Kovalev2018}.

On the other hand, skyrmions can be understood as stable, spatially localized solutions of
appropriate phenomenological continuum theories. 
In chiral magnets, the central theoretical framework for describing magnetic skyrmions
and other nontrivial magnetization textures was originally formulated by
Dzyaloshinskii~\cite{Dz64}. In its simplest isotropic form, the associated energy functional $W(\mathbf{m})$ comprises  exchange and Dzyaloshinskii--Moriya interactions (DMI), supplemented by a Zeeman coupling to an external magnetic field $\mathbf{H}$:
\begin{equation}
    W(\mathbf{m}) = \int_{\Omega} \left[A(\nabla \mathbf{m})^2 
+ D\,\mathbf{m} \cdot (\nabla \times \mathbf{m}) 
- \mu_0M\mathbf{m\cdot \mathbf{H}}\right] d^3 \mathbf{r}
\label{DzModel}
\end{equation}
In crystals lacking inversion symmetry, the DMI
\cite{Dz64,moriya} originates from spin--orbit coupling, which couples neighboring spins
via antisymmetric exchange and energetically favors twisted magnetic textures.
Crucially, chiral interactions fundamentally alter the scaling properties of the theory, thereby circumventing the Hobart--Derrick theorem~\cite{hobart,derrick} and stabilizing particle-like skyrmions against radial collapse.
The characteristic length scale of a magnetic skyrmion is then set by the balance between the exchange stiffness $A$ and the strength of the DMI $D$~\cite{Bogdanov94,Wiesendanger2016}.

Generally, within the continuum description of magnetic systems, chiral interactions are represented by inhomogeneous Lifshitz invariants (LI), i.e., antisymmetric terms linear in the first spatial derivatives of the magnetization~\cite{landau},
\begin{equation}
    \mathcal{L}^{(k)}_{i,j} = m_i \partial_k m_j - m_j \partial_k m_i \, .
    \label{LI}
\end{equation}
For cubic helimagnets belonging to the $23$ (T) crystallographic symmetry class—such as
MnSi~\cite{muhlbauer2009}, FeGe~\cite{FeGe}, and other B20 compounds, which are the focus of
the present work—the DMI reduces to a particularly simple isotropic form, as given in
Eq.~(\ref{DzModel}).

Bulk magnetic systems provide a genuinely three-dimensional (3D) environment for the
stabilization and manipulation of skyrmions. In such systems, axisymmetric skyrmionic
textures typically form extended filaments that align parallel to an applied magnetic
field or along easy axes imposed by magnetic anisotropies~\cite{leonov2019skyrmion}. Their stability is usually
confined to narrow regions of the phase diagram in the vicinity of the magnetic ordering
temperature, commonly referred to as precursor states
\cite{FeGe,pappas2009chiral,leonov2023precursor}.

By contrast, thin-film geometries offer a qualitatively different regime. Reduced
dimensionality, enhanced surface and interface effects, and the presence of effective
anisotropies substantially broaden the stability range of magnetic skyrmions, allowing
them to persist over wide intervals of temperatures and magnetic fields far beyond the
precursor regions characteristic of bulk materials~\cite{yu2011}.
For this reason, thin films and multilayers have become the primary experimental and
theoretical platforms for skyrmion research, enabling both robust stabilization and
controlled manipulation of individual skyrmions.

Nowadays, skyrmions are regarded as promising building blocks for the next generation of memory and logic devices. They are often described as being ``topologically protected'', possess nanometer-scale dimensions~\cite{Wiesendanger2016}, and can be efficiently manipulated by electric currents~\cite{jonietz2010spin}. A prominent example is the skyrmion racetrack concept~\cite{fert2013skyrmions}, in which information is encoded in isolated skyrmions that propagate along a narrow magnetic strip.

According to the commonly accepted paradigm~\cite{Bogdanov94,mukai2022}, skyrmion meta-matter forms via the condensation of isolated skyrmions through a nucleation-type phase transition in the sense introduced by de~Gennes~\cite{deGennes}. 
At a critical magnetic field in (\ref{DzModel}), the eigen-energy of an isolated 2D skyrmion becomes negative relative to the surrounding homogeneous state, triggering their  condensation into a periodic array with an equilibrium inter-skyrmion spacing, i.e., a hexagonal skyrmion crystal (SkX).
Upon increasing the magnetic field, the SkX transforms back into the homogeneous state through a continuous divergence of the lattice period at the same critical field, accompanied by the release of repulsively interacting isolated skyrmions \cite{Bogdanov94,kovacs2017mapping,mukai2022}.
An analogous scenario occurs in one-dimensional (1D) spiral states, albeit at different critical fields: repulsive isolated kinks condense into a chiral soliton lattice  once their eigen-energy becomes negative upon decreasing the field, whereas increasing the field causes the period to diverge and the lattice to disintegrate into isolated kinks \cite{togawa2012chiral}.

At the same time, not all chiral magnetic solitons conform to this nucleation--condensation paradigm. 
As demonstrated in Ref.~\cite{leonov2026skyrmionium}, a 2D lattice composed purely of skyrmioniums is intrinsically unstable within the Dzyaloshinskii model (\ref{DzModel}) and relaxes instead into a modulated one-dimensional spiral phase, which constitutes the true global energy minimum over the same parameter range. 
This instability can be traced to the structure of an isolated skyrmionium (also referred to as a \(2\pi\)-skyrmion, reflecting the doubled radial rotation of the magnetization from its center to the periphery \cite{Bogdanov99,Komineas,Nakamura}), which may be viewed as a circular realization of a chiral kink. 
Once the energy of a kink becomes negative, a chiral soliton lattice can tile space with unit efficiency, whereas a skyrmionium lattice necessarily leaves extended regions of homogeneous ``vacuum'' both inside the skyrmionium cores and between neighboring objects. 
This geometric inefficiency drives a deformation of the skyrmionium textures, leading to the collapse of their vacuum cores and their eventual elongation into a spiral state \cite{leonov2026skyrmionium}.

Chiral interactions of the same functional form as the Lifshitz invariants~(\ref{LI})
can arise in a wide range of physical systems beyond chiral magnets.
Notable examples include ferroelectrics with non-centrosymmetric parent paraelectric
phases, non-centrosymmetric superconductors, multiferroics~\cite{bogdanov2001,bode2007,wright1989},
and even metallic supercooled liquids and glasses~\cite{sethna1983}.
Localized topological states stabilized by such interactions are commonly referred to as
skyrmions, by analogy with the Skyrme model originally introduced to describe mesons and
baryons in nuclear physics~\cite{skyrme1961}.

CLCs provide an ideal model platform for investigating modulated structures on mesoscopic
length scales~\cite{ackerman2017}. At a fundamental level, chirality in liquid crystals
originates from the acentric shape of the constituent molecules, which gives rise to
chiral elastic couplings and ultimately underpins the formation of solitonic textures.
As a result, CLCs host a remarkably rich variety of naturally occurring and
laser-generated topologically nontrivial states, including hopfion textures characterized
by complex knotting and linking of the nematic director field~\cite{ackerman2017,ackerman2017b}.
Moreover, in CLCs the relevant material parameters can be continuously tuned over wide
ranges, experiments are typically performed under ambient conditions, and the resulting
textures can be directly visualized with high spatial resolution~\cite{ackerman2017,ackerman2017b}, frequently revealing
fine structural details that remain difficult to access in magnetic systems.
By contrast, the direct imaging of three-dimensional magnetization textures in
ferromagnets generally relies on sophisticated experimental probes, and only recently
have techniques such as electron holography and X-ray magnetic imaging enabled
experimental access to complex three-dimensional magnetic configurations~\cite{donnelly2020}.
Thus, CLCs may establish a natural conceptual bridge to ChMs, helping to reconcile closely
related solitonic textures that often appear under different names and are interpreted
within distinct physical frameworks in the two systems.

Solitons in CLCs are likewise predominantly investigated in thin-film geometries and are
confined within glass cells whose thickness is comparable to the intrinsic spiral
pitch~\cite{afghah2017,leonov2014,nych2017}. In contrast to magnetic systems, however,
liquid-crystal solitons are typically subject to strong boundary conditions in the form of
easy-axis anisotropy, commonly referred to as homeotropic surface anchoring
\cite{Oswald,kleman2003}. Such confinement profoundly reshapes the topological phase landscape:
(i) it removes continuous topological pathways that would otherwise connect metastable and
stable configurations, thereby enabling a systematic probing of the stability limits of
individual modulated phases;
(ii) it suppresses bulk instabilities, allowing enhanced control over soliton structure
and topology; and
(iii) it stabilizes solitonic textures that have no direct counterparts in unconfined bulk
systems.
Despite its clear physical relevance, however,  the role of surface-induced anisotropy has received comparatively little attention in ChMs within the magnetic skyrmion community.

In the present paper, we focus on two varieties of cholesteric fingers, commonly referred
to as fingers of the first and second types (CF--1 and CF--2)~\cite{gil1998surprising}.
Cholesteric fingers are among the earliest and best-studied localized modulated structures in chiral liquid crystals; experimentally, they appear as elongated, stripe-like textures embedded in an otherwise homogeneous or weakly twisted background (see, e.g., Refs.~\cite{meyer1976,de1995,Oswald}). 
Accordingly, our aim is not to reproduce or refine this extensive body of work, but rather to cast cholesteric fingers into a broader conceptual framework inspired by the magnetic skyrmion community, viewing them as solitonic objects and potential building blocks of solitonic meta-matter. 
From this perspective, CF--1 and CF--2  are reinterpreted as extended chiral solitons whose interactions, transformations, and possible condensation into ordered states can be analyzed using concepts and intuition developed in chiral magnetism.
In addition, our study is motivated by the observation that CF--1 and CF--2 can be viewed as two-dimensional cross-sections of Hopfions with Hopf indices \(Q_H=0\) and \(Q_H=\pm 1\), respectively.  
Recent work~\cite{hopfions} suggests that the metastability region of Hopfions is closely linked to the stability region of the corresponding cholesteric fingers, implying that Hopfions may be regarded as precursor states of these finger textures.

To pursue this goal, we adopt a unified phenomenological approach based on the
Dzyaloshinskii model~(\ref{DzModel}) and extend it to incorporate bulk uniaxial anisotropy,
such as that induced by an external electric field in CLCs, as well as surface anchoring
effects characteristic of confined liquid-crystal geometries.
In contrast to ChMs, the Zeeman coupling to an external magnetic field is absent in CLCs
and is therefore omitted from our analysis. Instead, a characteristic feature of CLCs is
that the role played by the magnetic field in chiral magnets is effectively replaced by
the combined action of bulk and surface anisotropies~\cite{leonov2021}.
While this minimal model does not aim to capture the full complexity of cholesteric liquid-crystal phase diagrams—which are known to depend sensitively on the anisotropy of the elastic constants \(K_1\), \(K_2\), and \(K_3\) within the Frank--Oseen phenomenology—it provides a controlled framework in which these constants are taken equal~\cite{Oswald}, allowing us to isolate the essential topological and geometric mechanisms underlying the formation of isolated cholesteric fingers and their ordered meta-matter.
The phenomenological model and the numerical algorithms used for minimizing the corresponding energy functional are introduced in the subsequent Section~\ref{sect:model}.

We begin our analysis by examining the internal structure of isolated cholesteric fingers
(Sec.~\ref{sect:structure}).
As a first step, we construct their axisymmetric counterparts by applying a global
rotation of the vector field~$\mathbf{m}$ by $90^\circ$ about a chosen coordinate axis.
This construction reveals that fingers of the second type map onto conventional
axisymmetric skyrmions, whereas fingers of the first type transform into so-called
droplets.
The latter can be interpreted as composite solitons consisting of two inequivalent
terminations: a rounded, skyrmion-like tip and a pointed, antiskyrmion-like tip
(Fig.~\ref{fig01}(g)).
Overall, the system admits two energetically degenerate realizations of CF--1 and two degenerate realizations of CF--2, which are related by  a $\pi$ rotation about the vertical axis.
We then analyze the key characteristics of the finger textures, including the spatial
distributions of the individual contributions to the total energy density
(Fig.~\ref{fig02}).
This analysis demonstrates that the emergence of topological features---such as a
nonvanishing topological charge density---is primarily induced by surface confinement
rather than by the bulk rotational DMI, which may even generate locally opposite twists in
regions adjacent to the confining surfaces.
Finally, the finger profiles reveal a predominantly repulsive interaction between fingers
embedded in the homogeneous background, whereas fingers surrounded by the conical phase
exhibit an effective attraction. The qualitative character of the finger--finger
interaction is independent of the finger type, although its quantitative features depend
on the specific finger combination.

In Sec.~\ref{sect:lattices}, we show that cholesteric fingers condense into extended
quasi-one-dimensional arrays in accordance with the principles introduced by
de~Gennes~\cite{de1995}, placing them in the same conceptual category as skyrmions and
spirals in chiral magnets discussed in the Introduction \ref{sect:intro}. The critical values of the
uniaxial anisotropy that drive the phase transitions between condensed and isolated
states, however, differ for CF--1 and CF--2.
Remarkably, the repulsive interaction between cholesteric fingers renders them genuinely
particle-like objects and enables their controlled mixing into rows with prescribed
periodicity. Such mixed arrays behave essentially identically to pure finger lattices of
the same overall periodicity and therefore represent a compelling platform for
information encoding in prototype devices. When embedded in the conical phase, however, cholesteric fingers develop an effective attractive interaction that is universal with respect to their variety and topological charge, and is quantified by the equilibrium inter-finger spacing and the depth of the interaction potential.

In Sec.~\ref{sect:thickness}, we extend our analysis to confined CLC layers with thicknesses
both smaller and larger than the intrinsic spiral pitch $\lambda$.
Both CF--1 and CF--2 lose their stability below a critical thickness, in agreement with
previous experimental and theoretical studies~\cite{ribiere1994optical,Oswald}.
For thicker samples, the critical fields marking the transition between isolated fingers
and condensed finger meta-matter gradually converge.
Interestingly, CF--2 admits two distinct size variants in sufficiently thick layers; these
variants can be reversibly selected by tuning the bulk uniaxial anisotropy.

Finally, we address the relation between isolated Hopfions and the stability regions of cholesteric fingers. 
Hopfions can be understood as torus-like three-dimensional solitons obtained by revolving a two-dimensional CF--1 or CF--2 texture about the \(z\)-axis. 
From this viewpoint, Hopfions naturally emerge as three-dimensional extensions of cholesteric fingers, and their existence and metastability are intimately connected to the parameter regime in which the corresponding finger textures are stable. 
This geometric correspondence provides a direct link between the stability of extended finger states and that of localized Hopfion solitons.

\section{Phenomenological model\label{sect:model}}

As outlined in the Introduction \ref{sect:intro}, the energy density of a noncentrosymmetric ferromagnet, including both bulk and surface contributions, can be written as a sum of exchange, DMI, and uniaxial anisotropy terms~\cite{leonov2014,leonov2021}:
\begin{align}
W(\mathbf{m}) = 
&\int_{\Omega} \left[
A(\nabla \mathbf{m})^2 
+ D\,\mathbf{m} \cdot (\nabla \times \mathbf{m}) 
- K_u (\mathbf{m} \cdot \mathbf{z})^2
\right] d^3 \mathbf{r}
\nonumber\\
&\quad
- \int_{\partial \Omega} K_s (\mathbf{m} \cdot \mathbf{z})^2 \, d^2 \mathbf{r}.
\label{functional}
\end{align}
$\Omega$ and $\partial \Omega$ are the magnet's volume and boundary, respectively.

The bulk energy terms in Eq.~(\ref{functional}) directly correspond to the elastic contributions of the Frank--Oseen free energy~\cite{Oswald} describing splay (\(K_1\)), twist (\(K_2\)), and bend (\(K_3\)) distortions of the director field, provided that the one-constant approximation \(K_1 = K_2 = K_3 = K\) is employed~\cite{kleman2003}. 
Within this approximation, the material parameters are related as
\(A \rightarrow K/2\) and \(D \rightarrow K q_0\), where \(q_0 = 2\pi/\lambda\) is the chiral wave number of the ground-state cholesteric phase and \(\lambda\) denotes the helicoidal pitch, i.e., the distance corresponding to one full \(2\pi\) rotation of the director \(\mathbf{m}\) along the helical axis. 
For many common CLCs, the elastic constants are of comparable magnitude, rendering the one-constant approximation a reasonable description of their ordinary elastic behavior.
%
%
%
The uniaxial anisotropy in CLC is typically induced by an applied electric field \(\mathbf{E}\) and takes an analogous functional form,
\(-(\varepsilon_0 \Delta\varepsilon/2)(\mathbf{m}\cdot\mathbf{E})^2\),
where \(\Delta\varepsilon\) is the dielectric anisotropy. 
We neglect spatial inhomogeneities of the induced anisotropy as well as higher-order surface contributions, such as the \(K_{13}\) or \(K_4\) terms~\cite{Oswald}.

%

To make the results general and directly applicable to any material system described by the functional~(\ref{functional}), we employ dimensionless units.
 The characteristic length scale is $L_D = A/D$, and the dimensionless anisotropy constant is defined as $k_u = K_u A / D^2$.
The system geometry corresponds to a thin film, infinite in the $x$- and $y$-directions, with a thickness $\nu$ along $z$ equal to one spiral pitch, $\lambda=4\pi L_D$. Both $K_u$ and the surface anchoring strength $K_s$ promote an easy-axis ferromagnetic state with magnetization aligned along $\mathbf{z}$. 

Surface anisotropy is conveniently quantified by the extrapolation length
\(\xi = A/K_s\), which provides a measure of how strongly the surface constrains the order parameter. 
Physically, \(\xi\) may be interpreted as the fictitious distance outside the sample at which the boundary condition becomes effectively rigid~\cite{tai2018}. 
The strong-anchoring regime corresponds to the limit \(\xi \to 0\), where the surface enforces a fixed orientation.
Introducing the helix pitch \(\lambda\) as the characteristic length scale, one obtains the dimensionless extrapolation length
\(\overline{\xi} = A / (K_s \lambda)\).
Alternatively, \(\overline{\xi}\) can be expressed as
\(\overline{\xi} = 1/(4\pi k_u^s \overline{t})\),
where \(k_u^s\) denotes the effective uniaxial anisotropy strength confined to a narrow surface layer of dimensionless thickness \(\overline{t}\).
In the following, we therefore parameterize the surface anchoring in terms of \(k_u^s\) and fix \(\overline{t} = 4\pi/N_z \approx 0.2\) throughout the manuscript. \(N_z = 64\) is the numerical grid size along the $z$ direction which  is fixed. The numerical grid size along the $x$ direction is fixed to $N_x = 1$, while the grid size along the $y$ direction, $N_y$, is varied depending on the specific numerical task (typically $N_y = 128$).

The energy minimization of the functional~(\ref{functional}) is carried out primarily using the GPU-accelerated code \textsc{mumax3} (version~3.10), which integrates the Landau--Lifshitz equation within a finite-difference framework to obtain relaxed magnetization configurations~\cite{mumax3}. 
The reliability of these numerical results is assessed independently by comparison with solutions obtained from in-house implementations of simulated annealing and single-spin Metropolis Monte Carlo algorithms. 
Details of these auxiliary methods have been reported previously~\cite{leonov2019skyrmion} and are therefore not reiterated here.

We emphasize that the model~(\ref{functional}) is employed here primarily as a mathematical and phenomenological framework. 
Accordingly, any direct comparison between the obtained solutions and experimental realizations in ChM or CLC must be made with due caution. 
In particular, the strong homeotropic surface anchoring that arises naturally in CLC is generally difficult to realize in magnetic systems. 
Although relatively strong effective surface anisotropies are known to occur at interfaces of strained chiral magnets~\cite{karhu2012chiral,huang2012extended}, at magnetic metal--oxide interfaces~\cite{dieny2017perpendicular}, in metallic multilayers~\cite{johnson1996magnetic}, and in chiral magnet--ferromagnet heterostructures~\cite{kawaguchi2016skyrmionic,porter2015manipulation}, their magnitude typically remains below the values characteristic of CLC and is likely insufficient to stabilize the solitonic textures considered here. 
Conversely, even a modest anisotropy of the Frank elastic constants in CLC can substantially modify the phase diagram, since stability is governed by the lowest energy scales in the hierarchy. 
Nevertheless, despite these material-specific limitations, the present results provide a controlled and conceptually transparent starting point for addressing solitonic phenomena in ChM and CLC on an equal theoretical footing.

\section{Internal structure of isolated cholesteric fingers CF--1 and CF--2}
\label{sect:structure}

We find it instructive to begin by harmonizing the terminology used in the ChM and CLC communities and by clarifying the inter-transformations between solitonic counterparts stabilized in different homogeneous backgrounds.

\subsection{Axisymmetric ``relatives'' of cholesteric fingers}

The chiral magnetic skyrmion shown in Fig.~\ref{fig01}(a) is an axisymmetric, vortex-like
soliton whose symmetry axis is oriented along the $x$ direction.
Here, we deliberately omit details of the specific model and parameter set required for
its stabilization and focus exclusively on its circular, particle-like structure.
For consistency with the geometry adopted for cholesteric fingers, the two-dimensional
skyrmion configuration is embedded in the $yz$ plane.

When the orientation of the surrounding homogeneous state changes from
$\mathbf{m}\parallel x$ to a direction perpendicular to this axis, e.g.,
$\mathbf{m}\parallel z$, the skyrmion continuously reshapes into a bound pair of coupled
merons that are structurally equivalent (Fig.~\ref{fig01}(b)).
By considering such a rotation, we deliberately abstract from any specific physical
mechanism that could realize this reorientation and instead focus on the resulting
topological and geometrical transformation of the texture.

The composite object shown in Fig.~\ref{fig01}(b) is commonly referred to as a
\emph{bimeron} in the skyrmion literature~\cite{lin2015,ohara2022}.
Each constituent meron carries a fractional topological charge \(Q=1/2\), with their axes aligned parallel and antiparallel to the original skyrmion axis, respectively. 
As a result, the bimeron preserves the total topological charge of the parent skyrmion while simultaneously adapting to the new orientation of the surrounding homogeneous state~\cite{toggle,duzgun2018}.

Experimentally and theoretically, magnetic bimerons have been considered in several distinct physical settings. 
They naturally arise in easy-plane chiral magnets~\cite{bachmann2023}, where in-plane anisotropy disfavors magnetization along the skyrmion axis and drives the splitting of a skyrmion into two meronic constituents \cite{mukai2024}. 
Alternatively, bimerons can be stabilized by applying an external magnetic field oblique to the skyrmion axis, with a finite angle approaching \(\pi/2\), which similarly enforces a reorientation of the background magnetization and induces the meron-pair configuration~\cite{barton2023}.

In the CLC community, topologically equivalent solitonic objects are known as cholesteric fingers of the second type, CF--2~\cite{Oswald}.
Their bimeron-like structure is enforced by strong homeotropic surface anchoring (Fig.~\ref{fig01}(c)).
Compared to the ideal bimeron shown in Fig.~\ref{fig01}(b), the internal structure of CF--2 is distorted because the two constituent merons carry opposite vorticities.
As a result, only the circular meron is energetically favored by the chiral interaction, whereas the other acquires a crescent-shaped, anti-meronic form. 

Numerical solutions for cholesteric fingers of the second type were obtained early on within continuum models employing the simplifying assumption of isotropic elasticity, see, e.g., Ref.~\cite{gil1998surprising}. 
Subsequent experimental studies focused on their remarkable robustness against applied electric fields and the associated nonlinear electro-optical response~\cite{ribiere1994optical}. 

In the CLC literature, CF textures are often analyzed using geometric representations in which the director field is mapped onto trajectories on the order-parameter sphere $S^2$, with each point on the sphere corresponding to a local director orientation~\cite{Oswald}. 
This construction provides a powerful qualitative tool for visualizing the internal structure of fingers, their stability, and their connection to the surrounding homogeneous state. 
In the present work, however, we deliberately refrain from employing such spherical representations and instead analyze cholesteric fingers directly within a unified field-theoretical framework that facilitates comparison with solitonic textures studied in chiral magnets.

By applying the same transformation to CF--2 in Fig.~\ref{fig01}(c)—namely, rotating all vectors by an angle of $90^\circ$ about the $y$ axis so that the homogeneous background is reoriented along the $x$ direction—we recover an ordinary skyrmion configuration, albeit with an elliptical deformation (Fig.~\ref{fig01}(d)).
We emphasize, however, that such rotations are not intended to establish a strict one-to-one correspondence between different solitons; rather, they are introduced to elucidate their underlying topology. 
Indeed, the solitons shown in Figs.~\ref{fig01}(b) and (d), obtained from their respective host configurations, exhibit slight distortions but remain topologically equivalent to those shown in Figs.~\ref{fig01}(a) and (c).
In the following, we  refer to bimerons as CF--2.

Conceptually similar transformations can be applied to the cholesteric finger of the first type, CF-1, shown in Fig.~\ref{fig01}(e). 
A rotation about the $y$ axis yields a composite soliton composed of two parts with opposite topological charges: a pointed segment with an anti-skyrmionic sense of rotation and negative vorticity, connected to a rounded tip with positive vorticity (Fig.~\ref{fig01}(f)). 
As a consistency check, we start from an ordinary skyrmion [Fig.~\ref{fig01}(a)] and reverse the vorticity in one half of the texture, which results in a topologically trivial droplet [Fig.~\ref{fig01}(g)]. 
Such two-dimensional droplets have been studied extensively in magnets~\cite{sisodia2021,zhou2015,rozsa2017} and can be stabilized not only by DMI or dipolar interactions, but also through dynamical nucleation and sustained by external driving.
A subsequent rotation of the spins within the droplet in Fig.~\ref{fig01}(g) reconstructs the CF--1 configuration [Fig.~\ref{fig01}(h)], completing the analogy. In the following, we refer to these solitons as CF--1. 
To our best knowledge, the first numerical investigations of CF--1 were carried out already in the 1970s and can be traced back to Ref.~\cite{press1976static}.

\begin{figure*}
  \centering
  \includegraphics[width=0.99\linewidth]{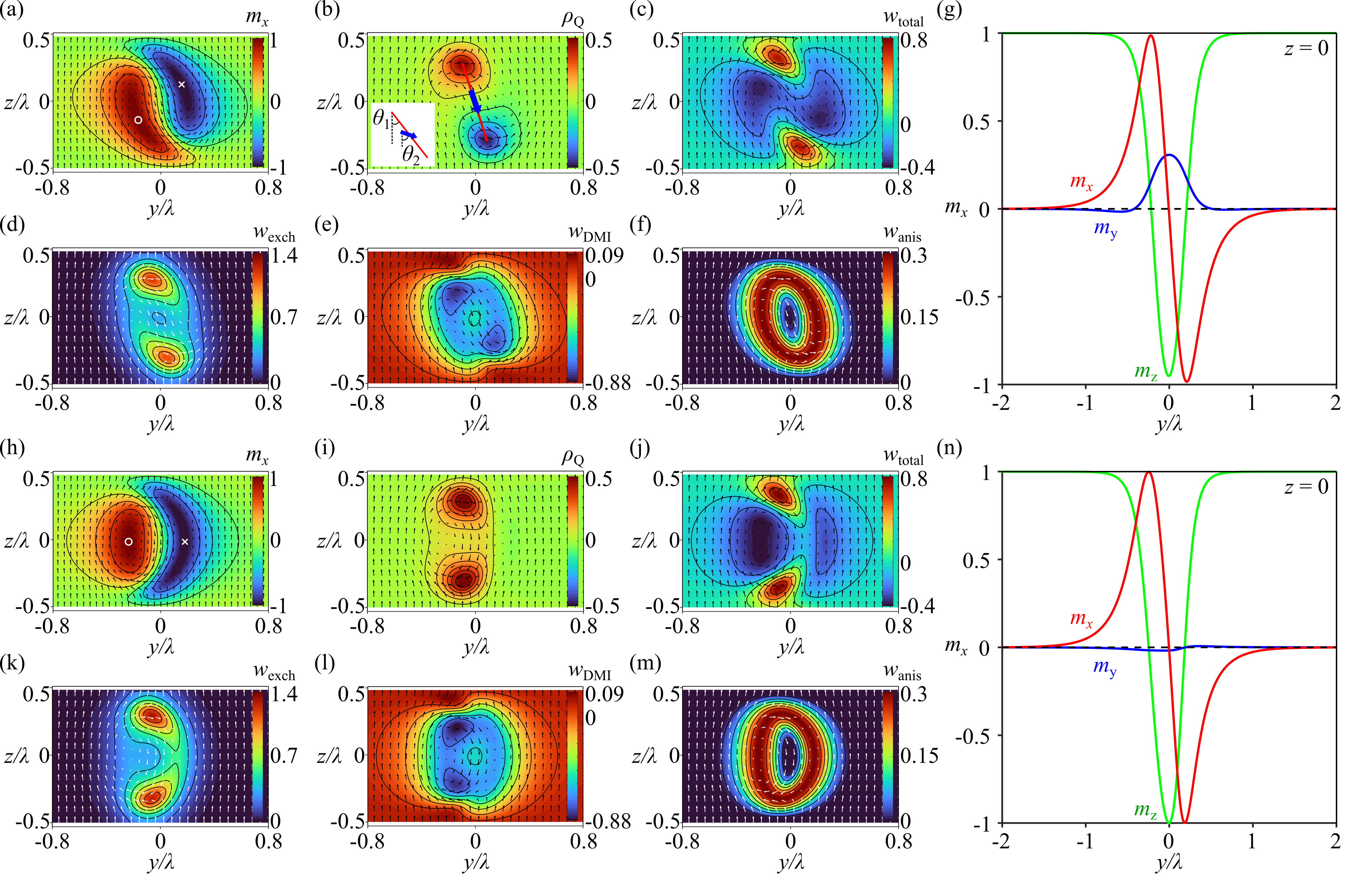}
  \caption{\label{fig02} Internal topological and energetic structure of isolated cholesteric
fingers CF--1 and CF--2 embedded in a homogeneous background at $k_u=0.33$ and
$k_u^s=40$.
(a),(h) Color plots of the magnetization component $m_x$ for CF--1 and CF--2,
respectively. The centers of the constituent merons are indicated by white dots and crosses.
(b),(i) Corresponding topological charge density distributions. In panel (b),
the characteristic angles $\theta_1$ and $\theta_2$ are indicated, which serve
as quantitative measures of the structural robustness of CF--1.
(c)--(f) and (j)--(m) Spatial distributions of the total, exchange, DMI, and anisotropy energy densities for CF--1 and CF--2.
Strong homeotropic surface anchoring induces pronounced confinement effects:
for CF--1, the meron centers are displaced toward opposite confining surfaces, leading to
antisymmetric meron profiles and an accumulation of topological charge density near the
surfaces.
In contrast, CF--2 exhibits mirror symmetry with respect to the midplane, with both meron centers located on the median plane.
As a consequence, CF--1 consists of two merons with identical vorticities and opposite
partial topological charges ($Q=0$), whereas CF--2 is composed of merons with opposite
vorticities whose partial charges add up to a total charge $Q=1$.  (g),(n) Magnetization profiles taken along the central planes of CF--1 and CF--2,
respectively.
Their exponentially decaying asymptotics indicate that cholesteric fingers embedded in a
homogeneous background interact repulsively, irrespective of the finger type [see text for details].
}
\end{figure*}

\subsection{Topological and energetic structure of CF solitons}

Figure~\ref{fig02} summarizes key internal characteristics of isolated CF--1 and CF--2
solitons embedded in a homogeneous background at $k_u = 0.33$ and $k_u^s = 40$
[see the corresponding phase diagrams in Fig.~\ref{fig05}(a),(b) for the location of this parameter point].

The color plot of the $m_x$ component within CF--1 [Fig.~\ref{fig02}(a)] closely resembles
the magnetization component perpendicular to the symmetry axis of an ordinary
axisymmetric skyrmion shown in Fig.~\ref{fig01}(a).
In particular, the corresponding $m_z$ component of the skyrmion would appear as two
joined half-disks of opposite sign.
In the case of cholesteric fingers, however, strong surface anchoring substantially
modifies the internal structure.
This confinement-induced distortion redistributes the magnetization across the film
thickness and enhances the robustness of CF--1 against collapse.
In particular, the meron centers are displaced toward opposite confining surfaces rather
than lying on the central plane, as indicated by the white dot and cross in
Fig.~\ref{fig02}(a).
As a consequence, each meron acquires an antisymmetric profile with a pointed tip on one side and a rounded tip on the other, so that the CF--1 deviates from a circular shape composed of two identical circular merons. 

The magnetization at the center of the structure is not strictly antiparallel to the
$z$ axis but instead forms a finite angle with it, denoted by $\theta_2$ and indicated in
Fig.~\ref{fig02}(b).
This angle provides a convenient quantitative measure of the internal structure of
CF--1 and will be used below as a structural parameter when analyzing thin films of
varying thickness in Sec.~\ref{sect:thickness}.
The energy density distributions shown in Figs.~\ref{fig02}(c)--(f) are all characterized
by a two-dimensional point-group symmetry $2$, with the rotation axis oriented along the
$x$ direction.
Interestingly, the DMI energy density in Fig.~\ref{fig02}(e) reveals the coexistence of two opposite senses of magnetization rotation. 
Regions with the unfavorable twist are confined to the vicinity of the surfaces and are induced by the strong homeotropic anchoring. 
As a result, the topological charge density [Fig.~\ref{fig02}(b)] accumulates near the
confining surfaces and becomes partially decoupled from the DMI energy density, in contrast to solitons in unbounded systems.
The rounded regions with positive and negative topological charge density are not located strictly above one another; instead, their centers are laterally displaced, forming a finite angle $\theta_1$.
This angle provides an additional quantitative measure of the confinement-induced
distortion and can be used to characterize the dependence of the internal finger structure on the film thickness in Sec.~\ref{sect:thickness}.

By contrast, the centers of the merons forming CF--2 [Fig.~\ref{fig02}(h)] are located on the
midplane of the film, and the magnetization in the region between the two merons points
exactly along the negative $z$ direction.
Accordingly, the energy density distributions shown in
Figs.~\ref{fig02}(j)--(m) exhibit mirror symmetry with respect to the central plane.
In particular, the rounded regions of topological charge density are aligned vertically and do not display any lateral displacement [Fig. \ref{fig02}(i)].

Thus, both CF--1 and CF--2 consist of two merons with opposite polarities. 
However, their vorticities differ: the two merons have the same vorticity in CF--1, whereas they have opposite vorticities in CF--2. 
As a result, the partial topological charges contribute with the same sign in CF--2 [Fig.~\ref{fig02}(i)] and with opposite signs in CF--1 [Fig.~\ref{fig02}(b)], yielding total topological charges $Q=1$ for CF--2 and $Q=0$ for CF--1.

Surface anchoring plays a crucial role in stabilizing cholesteric fingers. 
In the absence of surface anchoring, $k_u^s=0$, finger-like solutions continuously transform into spiral kinks whose central profiles extend homogeneously across the entire film thickness. 
This behavior provides a natural explanation for why, to the best of our knowledge, there is no experimental evidence for such finger textures in thin films of chiral magnets, where strong surface anchoring analogous to that in CLC is typically absent.

\begin{figure}
  \centering
  \includegraphics[width=0.99\linewidth]{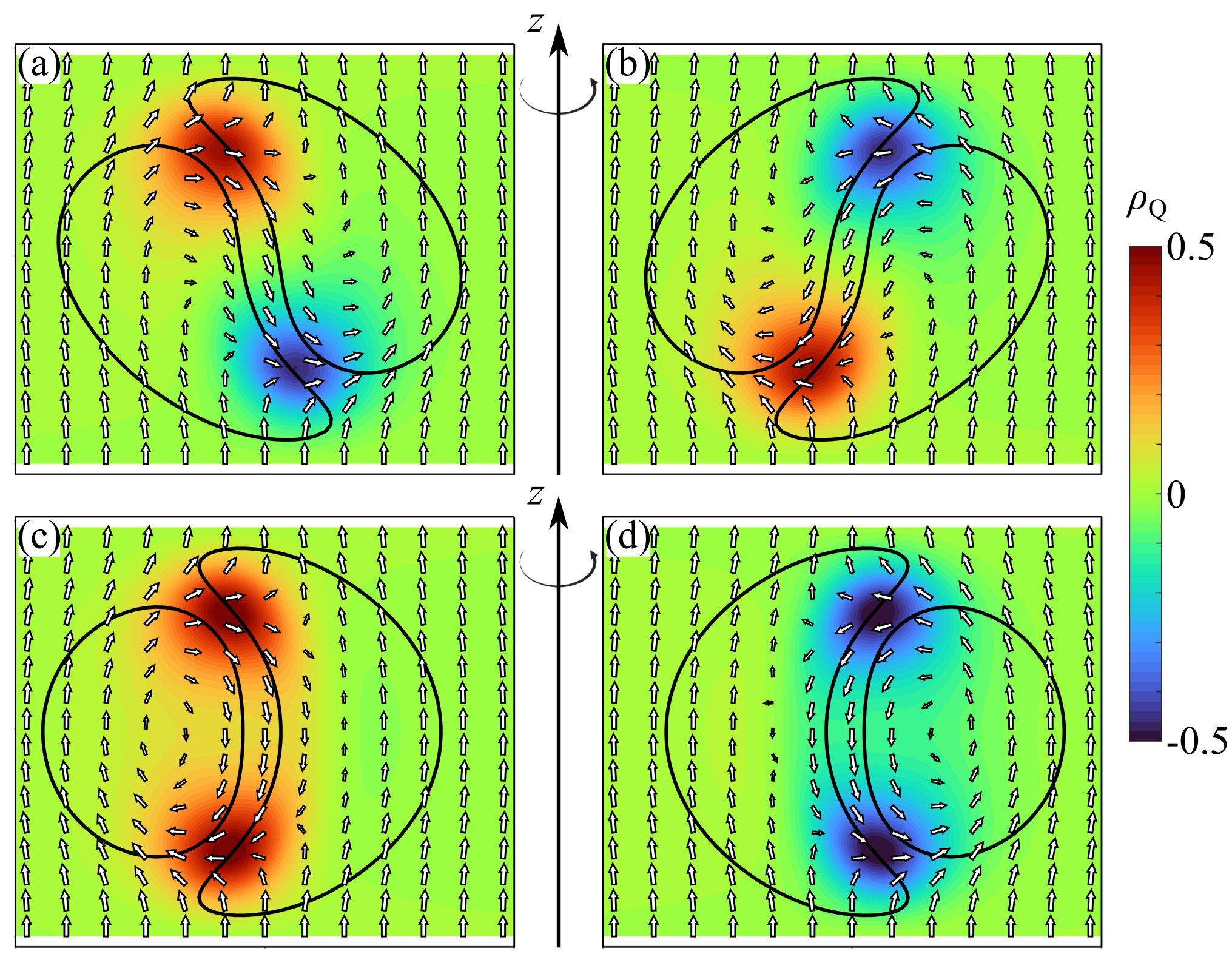}
  \caption{\label{fig03} Topologically and energetically equivalent realizations of cholesteric fingers of the first (CF--1) and second (CF--2) types with the correct rotational sense favored by DMI.
(a),(b) Two degenerate realizations of CF--1 related by a rotation of the texture by $\pi$ about the vertical $z$ axis, which reverses the polarity. Under this transformation, the pointed and rounded terminations of the composite meronic structure are exchanged, while the total energy and topology remain unchanged.
(c),(d) Two degenerate realizations of CF--2 obtained by the same $\pi$ rotation, corresponding to polarity reversal and an inversion of the total topological charge. Owing to mirror symmetry with respect to the film midplane, CF--2 admits no additional distinct realizations.
}
\end{figure}

\subsection{Distinguishable varieties of cholesteric fingers}

The topology of the cholesteric fingers considered in Fig.~\ref{fig02} implies the existence of several distinct, yet topologically and energetically equivalent, realizations of both CF--1 and CF--2.
For each finger type, only two realizations possess the rotational sense favored by the DMI and are therefore stable.
Other conceivable counterparts, obtained by reversing the sense of rotation, are
energetically unfavorable due to the wrong chirality and do not correspond to stable solutions.

Both stable finger counterparts are related by a rotation of the texture by $\pi$ about the vertical $z$ axis, which reverses the polarity, as illustrated in Fig.~\ref{fig03}(a)--(d). Under this transformation, the pointed and rounded terminations of the composite meronic structure forming CF--1 are exchanged [Fig.~\ref{fig03}(a),(b)], while the overall topology and total energy remain unchanged. In contrast, for CF--2 this rotation reverses the sign of the total topological charge [Fig. \ref{fig03}(c),(d)].

\begin{figure*}
  \centering
  \includegraphics[width=0.99\linewidth]{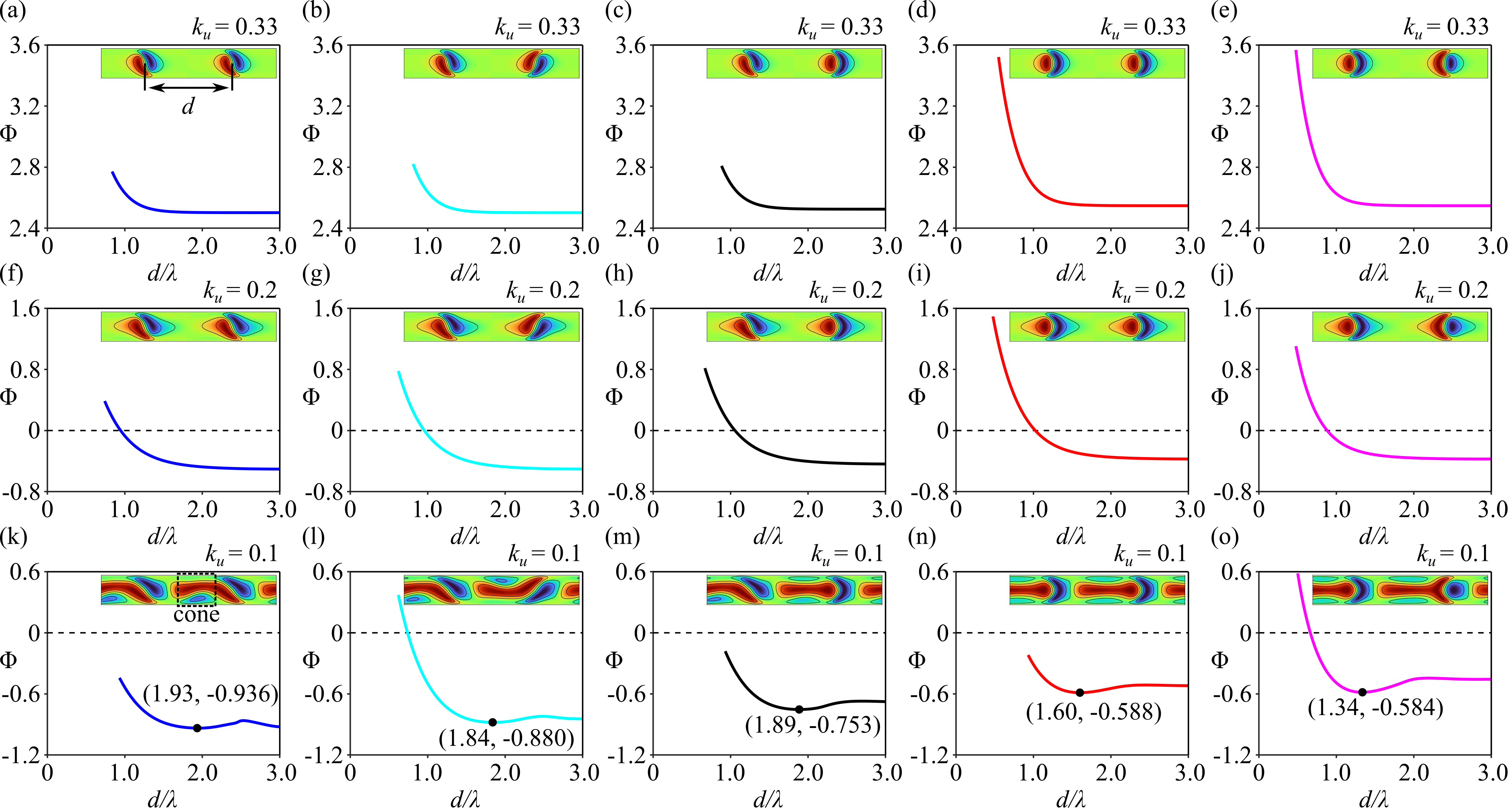}
  \caption{\label{fig04}
Interaction potentials between cholesteric fingers.
(a)--(e) Interaction potentials $\Phi(d)$ for identical and mixed pairs of CF--1 and CF--2 fingers embedded in the homogeneous state at $k_u^s=40$ and $k_u=0.33$. The total energy is shown as a function of the separation $d$ between the finger centers, measured with respect to the homogeneous state. All curves exhibit a monotonic decrease with increasing distance, demonstrating purely repulsive interactions independent of finger type or topological charge.
(f)--(j) Interaction potentials computed at parameters to the left of the zero--eigenvalue lines, where the eigen-energy of isolated fingers is negative. Despite this, the interaction remains repulsive, confirming that finger condensation into periodic phases is not driven by attractive forces.
(k)--(o) Interaction potentials for fingers embedded in the metastable conical background at $k_u^s=40$ and $k_u=0.1$. In this case, local minima develop at finite separation, indicating effective attractive interactions and the formation of equilibrium bound configurations.
Insets show representative color plots of the $m_x$ component for the corresponding pair configurations.
}
\end{figure*}

\subsection{Mutual repulsive interaction between cholesteric fingers surrounded by the homogeneous state}

The long-range behavior of the internal characteristics shown in Fig.~\ref{fig02}
unambiguously indicates that cholesteric fingers embedded in the homogeneous state interact
repulsively, irrespective of the finger type.
In particular, the magnetization profiles taken along the central planes of CF--1 and
CF--2 [Figs.~\ref{fig02}(g) and (n)] exhibit exponentially decaying asymptotics, which are
typical of isolated solitons stabilized in a homogeneous background.
Such behavior is well known for axisymmetric skyrmions [Fig.~\ref{fig01}(a)] as well as for
one-dimensional kinks and generally implies repulsive inter-soliton forces
\cite{leonov2016properties}.

This finding has important physical implications.
Despite their bimeron-like internal structure, both CF--1 and CF--2 behave as genuinely
particle-like objects whose mutual interaction is governed by repulsion.
This behavior stands in sharp contrast to magnetic bimerons in unbounded or weakly
confined systems, which typically exhibit attractive interaction potentials and a strong
tendency toward pairing or clustering~\cite{mukai2024,barton2023}.
In the present case, surface confinement and the associated deformation of the internal
structure suppress such attraction, rendering cholesteric fingers robust, mutually
repelling solitons.

To quantify the finger--finger repulsion, we compute the interaction potentials $\Phi$ for identical and mixed pairs of fingers, shown in Fig.~\ref{fig04} (first row, panels (a)--(e)), at the parameter point $k_u^s=40$ and $k_u=0.33$.
The interaction potential is obtained by evaluating the total energy~(\ref{functional}) for configurations in which the centers of two fingers are pinned at a prescribed separation $d$ [inset in Fig. \ref{fig04}(a)].
For clarity, the reference energy of two infinitely separated fingers is not subtracted; instead, the full total energy with respect to the homogeneous state is plotted as a function of $d$.
Insets show color plots of the $m_x$ component for representative paired configurations.

The qualitative behavior of all curves in Fig.~\ref{fig04}(a)-(c) is unambiguous.
None of the interaction potentials exhibits a local minimum at finite separation that
could stabilize bound states or clusters.
On the contrary, the energy decreases monotonically with increasing distance,
demonstrating that both identical and dissimilar cholesteric fingers prefer to remain well
separated.
Upon releasing the pinning constraint, the two bimeron-like objects would naturally move
apart while continuously readjusting the texture to minimize the total energy.
This confirms that CF--1 and CF--2 behave as mutually repelling, particle-like solitons
rather than forming composite or clustered states when embedded in the homogeneous phase.

Remarkably, pairs of CF--2 fingers with opposite topological charges [Fig. \ref{fig04}(e)] also exhibit a
repulsive interaction, which drives them to remain well separated rather than to
approach each other and mutually annihilate.

\begin{figure*}
  \centering
  \includegraphics[width=0.99\linewidth]{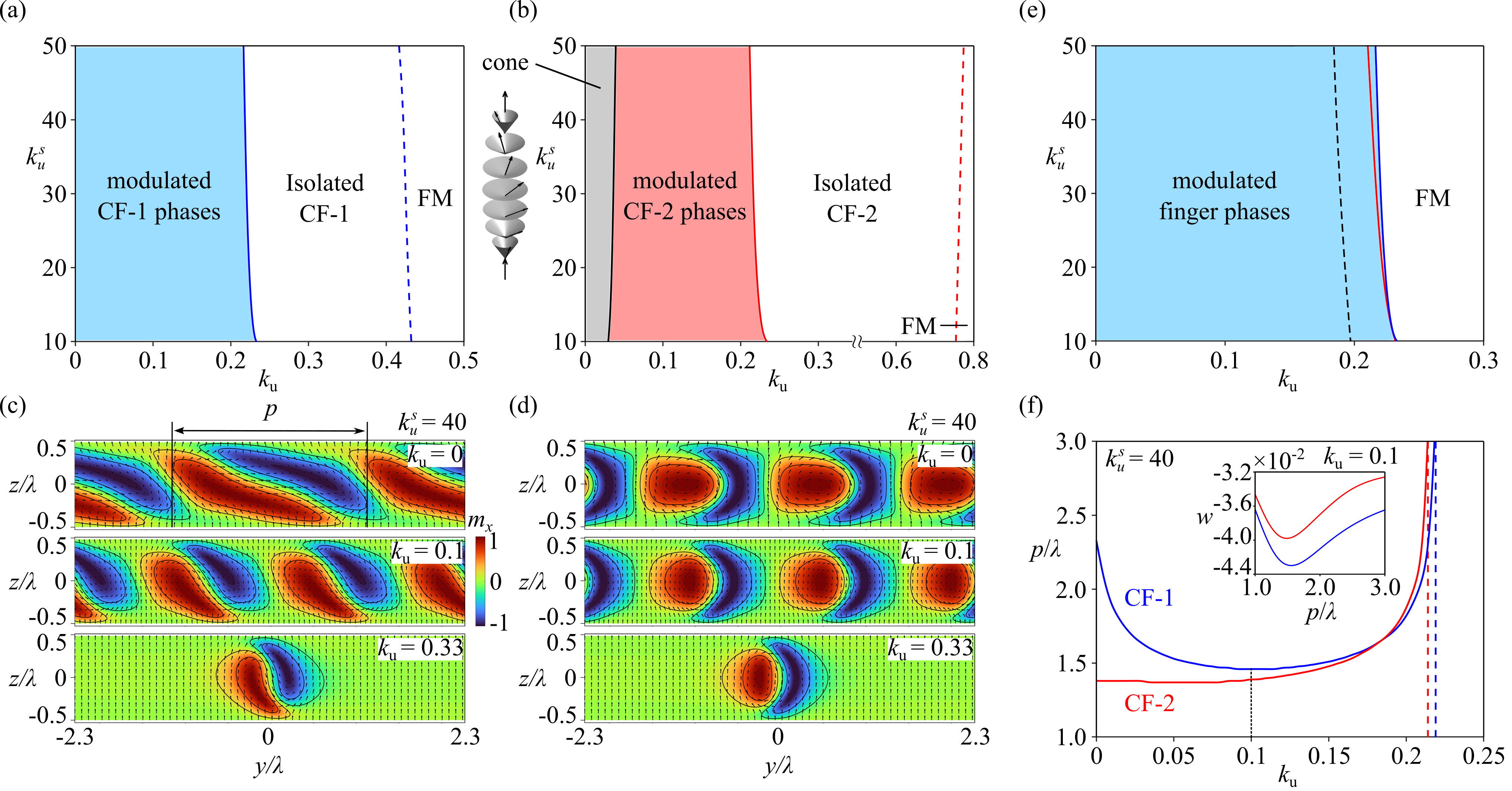}
  \caption{\label{fig05}
Phase diagrams and periodic finger phases.
(a),(b) Phase diagrams in the $(k_u^s,k_u)$ plane showing the stability regions of periodic CF--1 and CF--2 phases, respectively. Solid blue and red lines denote the zero–eigenenergy boundaries of isolated CF--1 and CF--2 fingers, equivalently marking the parameter lines where the equilibrium period of the corresponding modulated finger phases diverges. Dashed lines indicate the collapse of isolated fingers.
(c),(d) Representative spin configurations of periodic CF--1 and CF--2 phases at $k_u^s=40$ for different values of $k_u$ (top to bottom), together with examples of isolated fingers. Increasing uniaxial anisotropy renders CF--1 solitons more compact.
(e) Combined phase diagram highlighting the stability region of the CF--1 phase (blue shaded area) together with the expansion line of the CF--2 phase (red line). The mismatch between these critical lines underlies the formation of complex composite states discussed in the text. Black line indicates the phase boundary between the conical spiral and the homogeneous state.
(f) Equilibrium period $p$ of the CF--1 and CF--2 phases as a function of $k_u$ at fixed $k_u^s=40$, illustrating the divergence of the period at the corresponding stability boundaries. The inset shows the associated energy densities, demonstrating that the CF--2 phase has a higher energy density than the CF--1 phase. The presence of pronounced energy minima as functions of the period highlights the necessity of treating the lattice period as a variational parameter when determining stability regions of modulated phases.
}
\end{figure*}

\section{Modulated finger phases \label{sect:lattices}}

\subsection{Eigen-energy criteria and stability of CF--1 and CF--2 phases}

Despite their mutually repulsive interaction, cholesteric fingers can nevertheless
condense into periodic arrays once their eigen-energy becomes negative relative to the
homogeneous state.
The condensation process is therefore governed by a competition between two mechanisms:
the low-energy finger cores, which favor proliferation and space filling, and the
high-energy edge regions, which are responsible for the repulsive interaction between
isolated fingers.
As a consequence, the period of the finger phases is directly controlled by the
eigen-energy of an isolated finger. In particular, the period must diverge upon
approaching the parameter values at which the eigen-energy of a single finger becomes zero.
In this sense, the condensation of cholesteric fingers into their corresponding
meta-matter faithfully reproduces the line of arguments discussed in the Introduction \ref{sect:intro} for magnetic skyrmions and magnetic kinks and can be classified, following de~Gennes, as a nucleation-type phase transition \cite{deGennes}.

Figures~\ref{fig05}(a) and (b) show the zero–eigenvalue lines for CF--1 and CF--2 (blue and red solid lines, respectively) on the phase diagram plotted in the $(k_u^s,k_u)$ plane.
Within the blue- and red-shaded regions, CF--1 and CF--2 exist as periodic structures with a finite characteristic period~$p$, which diverges at the boundaries of the corresponding stability domains.
This divergence is illustrated in Fig.~\ref{fig05}(f), where the period is plotted as a function of the uniaxial anisotropy~$k_u$ at a fixed surface anisotropy $k_u^s=40$.

For the CF--2 phase, the period increases monotonically from its minimal value at $k_u=0$ and diverges at the upper boundary of the red-shaded region [Fig. \ref{fig05}(f)].
In contrast, the CF--1 phase exhibits a nonmonotonic behavior: its period initially
decreases with increasing $k_u$, reaches a minimum near the center of the blue-shaded
region, and then increases again, diverging at the phase boundary [Fig. \ref{fig05}(f)].
We note that the period $p$ of the CF--1 phase remains finite, although rather large, at $k_u=0$, which renders individual CF--1 solitons spatially extended [upper panel in Fig.~\ref{fig05}(c)].
In this sense, the uniaxial anisotropy $k_u$ provides an efficient control parameter for rendering CF--1 solitons more compact [middle panel in Fig.~\ref{fig05}(c)].

The inset of Fig.~\ref{fig05}(f) demonstrates that, for each parameter point within the
stability regions, the equilibrium period $p$ of the finger phases is obtained by
minimizing the total energy density defined as 
\begin{equation}
    \varepsilon = \frac{1}{p\lambda} \int w(x,z)\, dxdz.
    \label{epsilon}
\end{equation}
The color plots in Figs.~\ref{fig05}(c) and (d) illustrate representative spin
configurations of the equilibrium CF--1 and CF--2 phases, together with examples of isolated fingers [bottom panels in Fig. \ref{fig05}(c),(d)].
This methodological point is particularly important, since in many studies of solitons in CLCs the lattice period is not treated as a variational parameter. 
As a result, incorrect conclusions regarding the stability of specific phases and even erroneous phase boundaries may be drawn, as exemplified by Ref.~\cite{tai2018}.

To further demonstrate that the interaction remains repulsive also to the left of the critical lines in Fig.~\ref{fig05}(e), and thereby substantiate the arguments presented above, we compute the interaction potentials at
$k_u=0.2$ and $k_u^s=40$ (second row in Fig.~\ref{fig04}, panels (f)-(j)).
Despite the negative eigen-energy with respect to the homogeneous state, the fingers still prefer to avoid each other, and no attractive minimum emerges in the interaction potential.
Only by forcing a higher finger density—i.e., by packing the system with an increasing number of fingers—does the system reach the equilibrium period $p$ of the condensed phase.

\begin{figure*}
  \centering
  \includegraphics[width=0.99\linewidth]{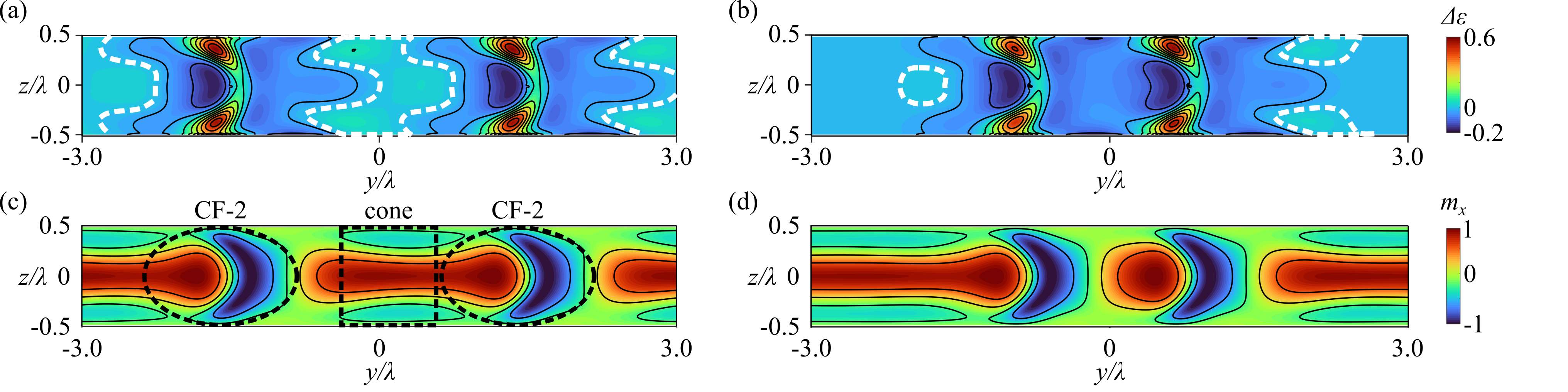}
  \caption{\label{fig06}
Energy-density mechanism underlying attractive interactions between cholesteric fingers CF-2 in the conical background.
(a) Energy density distribution for a remote CF2--CF2 pair embedded in the conical phase. The energy density is computed with respect to the conical spiral and reveals extended regions of positive excess energy surrounding each finger.
(b) Energy density distribution for a clustered CF2--CF2 configuration at the equilibrium separation. Upon approaching each other, the overlap of the distortion regions suppresses the inter-finger excess energy, resulting in an effective attractive interaction and stabilizing a bound state at a finite distance.
(c),(d) Corresponding color plots of the $m_x$ component of the magnetization.
}
\end{figure*}

The combined phase diagram in Fig.~\ref{fig05}(e) highlights the stability region of the CF--1 phase together with the expansion line of the CF--2 phase (red line). We note that the energy of the CF--2 phase is higher than that of the CF--1 phase, as also illustrated by the inset in Fig.~\ref{fig05}(f).
As shown in Ref.~\cite{hopfions}, the mismatch between these critical lines gives rise to nontrivial phenomena, most notably the formation of so-called \emph{hopfion bags}.
Along the red line in Fig.~\ref{fig05}(e), the energy of a circular CF--2—which is
topologically equivalent to a hopfion with Hopf index $Q_H=1$—approaches zero while its
characteristic size diverges.
As a consequence, the interior of such a hopfion can accommodate other solitonic states,
including the CF--1 phase, which is already stable in this parameter regime.
In this sense, the hopfion forms a ``cell'' whose boundary has vanishing energy cost and
exerts no pressure on its interior content.
At the blue line, the energy of the circular CF--1—which can be viewed as a hopfion-like
structure with Hopf index $Q_H=0$—also approaches zero.
This implies that it may form a cell-like structure whose interior can be filled with
other solitons, such as other hopfions, CF--2 textures, and/or torons, which are effectively
shielded from the surrounding environment \cite{hopfions}.

We emphasize that the phase diagram shown in Fig.~\ref{fig05}(e) is not a new result of the
present work, as it was already reported in Ref.~\cite{ribiere1994optical}, albeit in a
different representation (see also the up-to-date review in Ref.~\cite{Oswald}).
Rather than reproducing known phase boundaries, we focus on more subtle features of the phase diagram that underpin the main idea of the present paper. 
In particular, to the best of our knowledge, there exists no experimental study of CF--1 and CF--2 phases that traces the continuous increase of their spatial period up to divergence. 
This stands in marked contrast to the classical experimental verification of the Dzyaloshinskii theory in chiral magnets, where the chiral soliton lattice releases isolated kinks as the magnetic field increases, leading to a diverging period, as demonstrated in Ref.~\cite{togawa2012chiral}. 

In addition, the phase diagrams shown in Figs.~\ref{fig05}(a) and (b) also delineate the
collapse boundaries of isolated cholesteric fingers, indicated by dashed blue and red lines for CF--1 and CF--2, respectively.
Crossing these lines corresponds to the loss of metastability of an individual finger, which collapses into the homogeneous state under increasing uniaxial anisotropy.
A comparison of the two collapse boundaries demonstrates that CF--2 solitons are
significantly more robust against the destabilizing action of the uniaxial anisotropy.

\subsection{Mutual attractive interaction between cholesteric fingers surrounded by the conical state}

Interestingly, although the CF--2 phase remains a local energy minimum within the gray-shaded region of Fig.~\ref{fig05}(b), the conical state has a lower energy there and thus becomes the thermodynamically stable phase. 
In contrast, the CF--1 phase remains energetically favorable compared to the conical and CF-2 state throughout the entire range of the uniaxial anisotropy $k_u$ shown in Fig.~\ref{fig05}(a).

In the CLC literature, the conical phase is usually referred to as the \emph{Translationally Invariant Configuration} (TIC) phase~\cite{Oswald}. 
The TIC state is characterized by a helicoidal modulation of the director field with the helical axis oriented perpendicular to the confining substrates, while the director itself is uniformly tilted with respect to this axis, resulting in a finite projection along the surface normal. 
An important distinction from the conical phase in ChM is that, in the TIC phase, the polar angle of the director is generally not constant across the film thickness: it varies along the confinement direction and typically approaches zero at the surfaces due to homeotropic anchoring [inset in Fig. \ref{fig05}(b)]. 
By contrast, in bulk chiral magnets the cone angle remains uniform throughout the sample and can deviate from its equilibrium value only weakly, for example due to cubic or exchange anisotropies (see, e.g., Ref.~\cite{leonov2026low}). 
Uniaxial anisotropy in CLCs (as well as an external magnetic field in ChMs, which effectively closes the cone) provides an efficient mechanism for suppressing the conical/TIC phase.

In the absence of surface anchoring, the conical phase admits an analytical solution of the Euler--Lagrange equations derived from Eq.~(\ref{DzModel}):
\begin{equation}
    \psi = \frac{2\pi z}{\lambda}, \qquad 
    \theta = \frac{2h}{1 - 4k_u},
\end{equation}
where $\psi$ and $\theta$ denote the azimuthal and polar angles of the magnetization, respectively. \(h = \mu_0 M_s H A / D^2\) is the non-dimensional value of the applied magnetic field. 
For vanishing surface anisotropy, $k_u^s=0$, the conical phase becomes unstable at the critical uniaxial anisotropy $k_u = 0.25$.
In the following, we refer to this state simply as the conical phase.
In bulk chiral magnets, in particular, the conical phase is widely recognized as the principal competitor of skyrmions, as it severely restricts their stability to the narrow A-phase pocket \cite{muhlbauer2009,Crisanti}.
Consequently, a variety of strategies aimed at suppressing the conical phase are currently
under active development~\cite{leonov2026low}.

As shown in the phase diagram in Fig.~\ref{fig05}(e), the conical phase saturates at smaller values of $k_u$ than the CF--1 and CF--2 phases.
Despite being metastable, the conical phase renders the interaction between cholesteric fingers attractive.
Accordingly, the interaction potentials develop local minima corresponding to finite equilibrium inter-finger separations [third row in Fig.~\ref{fig04}, panels (k)--(o)].
The physical origin of these minima is analogous to that of isolated skyrmions embedded in a conical background (see Ref.~\cite{metamorphoses} for details).
In essence, in order to match the surrounding conical state, an isolated soliton acquires additional spatial twists associated with a positive excess energy density.
When two solitons approach each other, the overlap of these distortion regions reduces the total energy, resulting in an effective attractive interaction; see, e.g., Ref.~\cite{loudon2018direct} for a detailed discussion.
Figure~\ref{fig06} illustrates this mechanism quantitatively by showing the energy density distributions for distant and clustered CF--2--CF--2 pairs.
The energy density, computed with respect to the conical spiral, exhibits an extended region of positive excess energy surrounding both fingers when they are well separated [Fig.~\ref{fig06}(a)].
When the fingers merge into a bound configuration [Fig.~\ref{fig06}(b)], this inter-finger region of positive energy density is largely eliminated, thereby explaining the emergence of an energy minimum at a finite inter-finger distance~$d$.

The interaction potentials shown in the third row of Fig.~\ref{fig04} for
$k_u = 0.1$ and $k_u^s = 40$ exhibit a remarkable diversity in both the equilibrium inter-finger distances and the depths of the corresponding potential wells, depending on the specific combination of interacting fingers.


\begin{figure*}
  \centering
  \includegraphics[width=0.99\linewidth]{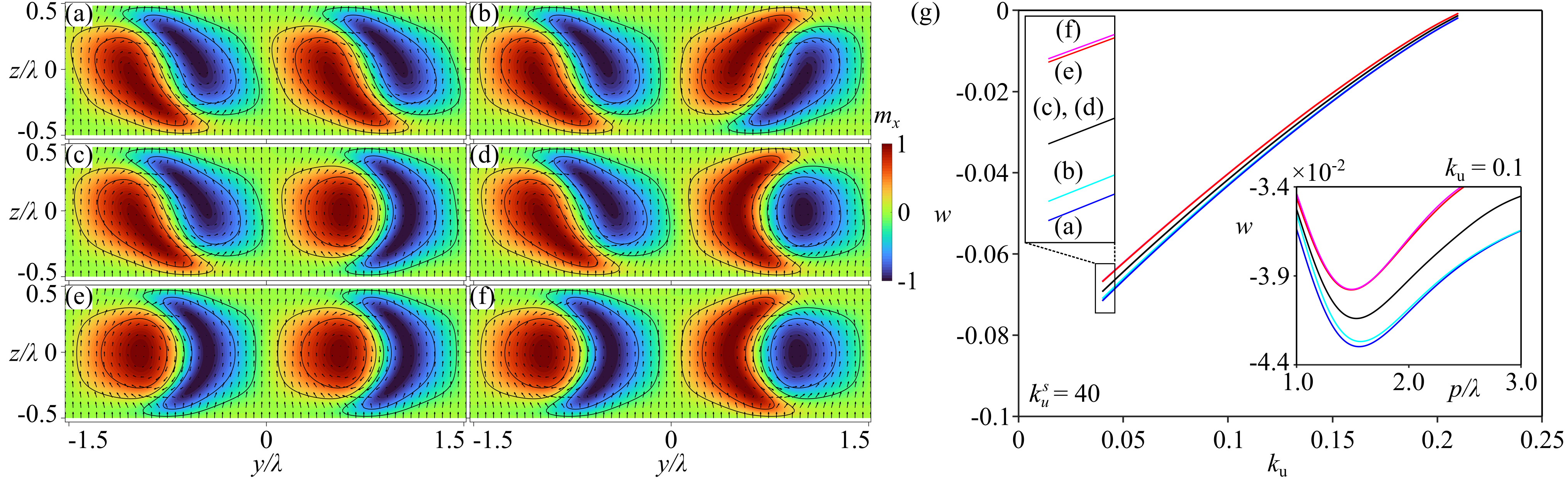}
  \caption{\label{fig07}
Mixed periodic states of cholesteric fingers and their energetic hierarchy.
(a)--(f) Color plots of the $m_x$ component for representative mixed modulated states composed of two different finger types within the unit cell, together with the pure periodic CF--1 state for comparison.
(g) Energy density of mixed finger phases as a function of finger composition, illustrating an effective ``energy band'' formed by different combinations of CF--1 and CF--2 fingers.
The lowest energy corresponds to the pure CF--1 phase, while the energy increases systematically with the number of CF--2 fingers in the unit cell, reaching a maximum for configurations composed exclusively of CF--2 fingers with opposite topological charges.
The inset shows the energy minimization with respect to the period for each mixed state, while the second inset provides a magnified view of the energy spectrum highlighting the fine energetic splittings between closely competing configurations.
}
\end{figure*}

\subsection{Band structure of mixed states of cholesteric fingers} 

The multiplicity of cholesteric fingers also enables the formation of mixed modulated states composed of a prescribed—or even random—succession of different finger types.
Color plots of the $m_x$ component for representative mixed states with characteristic periods comprising two distinct finger types are shown in Fig.~\ref{fig07}(a)--(f).

Among these configurations, the lowest energy density corresponds to the pure periodic CF--1 phase [Fig.~\ref{fig07}(a)].
Mixed states occupy higher positions in the energy hierarchy [Fig.~\ref{fig07}(g)], with the energy density increasing systematically with the number of CF--2 fingers present within the unit cell.
The highest energy density is obtained for mixed states composed exclusively of CF--2 fingers with opposite topological charges [Fig.~\ref{fig07}(f)].

In this sense, the ensemble of finger textures can be viewed as forming an effective ``energy band'' composed of several distinct finger varieties [Fig.~\ref{fig07}(g)].
Within this band, different finger types can be combined into periodic sequences at a moderate energetic cost, giving rise to a broad family of metastable mixed textures.
The inset in Fig.~\ref{fig07}(g) demonstrates that each mixed finger configuration is minimized with respect to its period.
A second inset provides a magnified view of the energy spectrum, resolving the fine energetic splittings between closely competing mixed states.

This perspective naturally motivates a combinatorial analysis, which allows one to estimate the number of distinct periodic finger configurations compatible with a given set of elementary fingers.

\subsection{Combinatorics of periodic finger arrangements}

The coexistence of several energetically degenerate varieties of cholesteric fingers naturally gives rise to a large number of distinct periodic finger arrangements.
In the present system, four elementary building blocks are available: two polarity-related realizations of CF--1 and two polarity-related realizations of CF--2, all possessing the correct rotational sense favored by the DMI.
Upon imposing periodic boundary conditions, a periodic finger phase can therefore be represented as a one-dimensional sequence of length $p$ drawn from an alphabet of four symbols, with sequences related by cyclic permutations describing the same physical structure.

Mathematically, such periodic arrangements are classified as \emph{necklaces} of length $p$ over an alphabet of four symbols.
The total number $N_p$ of inequivalent periodic finger structures with $p$ fingers per period is given by Pólya’s enumeration theorem as
\begin{equation}
N_p = \frac{1}{p}\sum_{d\,|\,p} \varphi(d)\,4^{p/d},
\label{eq:necklace}
\end{equation}
where the sum runs over all positive divisors $d$ of $p$ and $\varphi(d)$ denotes Euler’s totient function.

\paragraph*{Example: $p=2$.}
For two fingers per period, Eq.~(\ref{eq:necklace}) yields
\begin{equation}
N_2 = \frac{1}{2}\left(4^2 + 4\right) = 10.
\end{equation}
Among these configurations, four are non-primitive and correspond to uniform sequences with minimal period one.
The remaining six necklaces represent genuinely period--2 mixed states composed of two distinct finger varieties, including combinations of CF--1 and CF--2.
Representative examples of such mixed states are shown in Fig.~\ref{fig07}(a)--(f).

\paragraph*{Example: $p=3$.}
For three fingers per period, one finds
\begin{equation}
N_3 = \frac{1}{3}\left(4^3 + 2\times4\right) = 24.
\end{equation}
Again, four configurations correspond to uniform states, while the remaining $20$ necklaces describe genuinely period--3 structures.
These include sequences composed of two identical fingers and one different finger, as well as arrangements built from three distinct finger varieties.

More generally, the number of \emph{primitive} periodic structures—i.e., those whose minimal period is exactly $p$—is obtained by Möbius inversion as
\begin{equation}
N_p^{\mathrm{prim}} = \frac{1}{p}\sum_{d\,|\,p} \mu(d)\,4^{p/d},
\end{equation}
where $\mu(d)$ denotes the Möbius function.
The rapid growth of $N_p^{\mathrm{prim}}$ with increasing $p$ highlights the pronounced combinatorial richness of mixed finger phases.
This observation supports the existence of a large manifold of metastable periodic states even before energetic selection mechanisms are taken into account.

\subsection{Jagodzinski-type classification of finger sequences}

Beyond a purely combinatorial enumeration, periodic arrangements of cholesteric fingers
can also be classified using a local-environment perspective that is closely analogous to
the description of stacking sequences in close-packed crystals.
In crystallography, stacking variants of close-packed layers are commonly analyzed using
the nomenclature introduced by Jagodzinski~\cite{jagodzinski1949eindimensionale}, which plays a central
role in the theory of polytypism in metals and semiconductors.

In this framework, each close-packed layer is labeled according to its immediate
neighbors: a layer surrounded by two \emph{different} layers is assigned the symbol
\emph{c} (for cubic), while a layer surrounded by two \emph{identical} layers is assigned
the symbol \emph{h} (for hexagonal).
For example, the ideal fcc stacking sequence $\mathrm{ABCABC}\ldots$ corresponds to a
$\mathrm{cccc}\ldots$ sequence, whereas the hcp stacking
$\mathrm{ABAB}\ldots$ is described by $\mathrm{hhhh}\ldots$.
More complex polytypes, such as $\mathrm{ABAC}$ or $\mathrm{ABCACB}$, give rise to mixed
$\mathrm{h}$–$\mathrm{c}$ sequences that encode local stacking faults and domain
structures~\cite{verma1967polymorphism}.

An analogous classification can be introduced for periodic finger phases.
Here, a one-dimensional sequence of cholesteric fingers plays the role of the stacking
sequence, while the \emph{local environment} of a given finger is determined by the types
of its nearest neighbors along the periodic direction.
If a finger is flanked by two identical finger types, it may be labeled \emph{h}, whereas
a finger located between two different neighbors is labeled \emph{c}.
This Jagodzinski-type labeling provides a compact description of the local order within a
periodic finger phase, independent of the absolute length of the period.

Within this viewpoint, uniform finger phases correspond to purely \emph{h}-type sequences,
while strongly mixed phases are dominated by \emph{c}-type environments.
Intermediate cases, characterized by alternating \emph{h} and \emph{c} segments, represent
finger analogues of polytypic stacking variants in close-packed metals.
Importantly, many distinct finger sequences that are combinatorially inequivalent may
share the same Jagodzinski signature, indicating that they possess identical local
environments but differ in their global arrangement.

This local classification complements the global necklace enumeration discussed above and
offers a physically transparent way to analyze finger meta-matter in terms of local
packing rules, stacking faults, and domain structures—concepts that are deeply ingrained
in the theory of close-packed crystal structures.

\subsection{Prospects for finger-based spintronic schemes}

The cholesteric-finger textures discussed here could act as particle-like information carriers in spintronic systems, provided that analogous solitons can be realized in two-dimensional chiral magnets with sufficiently strong surface anisotropy.
In this regime, the system naturally supports four distinct and topologically robust solitons (two CF--1 and two CF--2 varieties), forming a finite set of discrete states suitable for multi-level information encoding and thus extending the binary paradigm.

Beyond isolated fingers, the ability to form periodic, mixed, and even disordered sequences of different finger types vastly enlarges the space of accessible textures.
As shown above, already simple periodic arrangements generate a combinatorially large number of inequivalent configurations, giving rise to a dense manifold of metastable states.
This situation closely parallels polytypism in close-packed crystals and suggests information-processing concepts based on sequence selection rather than on individual soliton manipulation.

A further advantage of finger-based textures stems from their topology.
In particular, CF--1 solitons carry zero total topological charge, which suppresses the skyrmion Hall effect and enables straight-line motion under applied spin torques.
Such behavior is highly desirable for racetrack-type device architectures, where transverse deflection limits controllability.

Taken together, the topology, degeneracy, interactions, and combinatorial richness of cholesteric fingers reveal a class of soliton phenomena that is presently far more developed in chiral liquid crystals than in magnetic systems.
This contrast points to a clear materials-science opportunity: engineering and enhancing surface-induced anisotropies in chiral magnets to levels comparable to those intrinsic to CLCs.
Achieving this would open access to an extensive family of finger-based magnetic solitons and their associated meta-matter, enabling spintronic functionalities rooted in confinement, topology, and structural diversity.

\begin{figure*}
  \centering
  \includegraphics[width=0.99\linewidth]{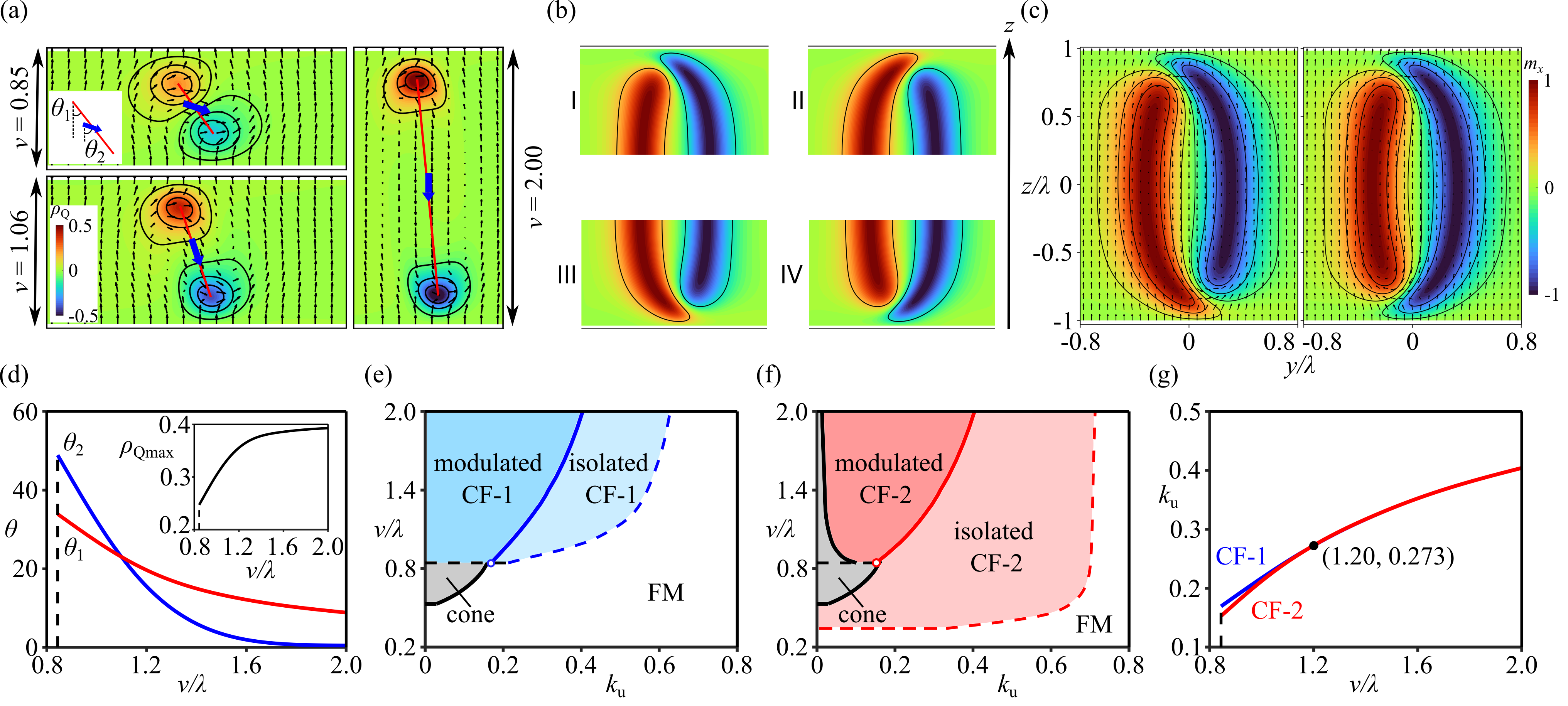}
  \caption{\label{fig08}
Thickness dependence of cholesteric fingers and stability limits of their modulated phases.
(a) Color plots of the topological charge density for CF--1 at three representative reduced film thicknesses, $\nu=0.85$, $1.06$, and $2$, illustrating the progressive tilting of the magnetization between the charge accumulations and their compression toward the central plane as the film becomes thinner.
(b) Upper and lower halves of CF--1 and CF--2 obtained by cutting the fingers by the central plane of the film. In the thick-film regime the upper halves become structurally indistinguishable.
(c) Examples of finger textures obtained by combining the upper and lower halves shown in (b), yielding four possible configurations.
(d) Thickness dependencies of the structural angles $\theta_1(\nu)$ and $\theta_2(\nu)$ for CF--1. The inset shows the maximal topological charge density, which vanishes at the critical thickness.
(e) Phase diagram for CF--1 at fixed $k_u^s=40$. The solid blue line denotes the zero--eigenvalue line of isolated CF--1 fingers, corresponding to the expansion boundary of the CF--1 phase, while the dashed blue line indicates the collapse of isolated fingers. The tricritical point at $\nu\approx0.8$ marks the disappearance of CF--1 where the conical and homogeneous phases also meet.
(f) Phase diagram for the CF--2 phase. The periodic CF--2 lattice exists to the left of the solid red line, while the gray region corresponds to the conical phase. The red shaded region bounded by the dashed red line indicates the metastability domain of isolated bimerons beyond the instability of the periodic lattice.
(g) Evolution of the critical lines with film thickness. Above $\nu\approx1.2$, the CF--1 and CF--2 instability lines merge, reflecting the increasing similarity of their internal structures.
}
\end{figure*}

\section{Effect of film thickness on the structure of cholesteric fingers \label{sect:thickness}}

\subsection{Stability limits of isolated fingers and their modulated phases}

\subsubsection{Effect of decreasing film thickness}

Upon decreasing the film thickness, both CF--1 and CF--2 eventually lose stability and disappear.

For CF--1, this instability is accompanied by a systematic increase of the structural angles $\theta_1$ and $\theta_2$. In Fig.~\ref{fig08}(a), color plots of the topological charge density are shown for three characteristic film thicknesses, $\nu=0.85$, $1.06$, and $2$. It is seen that, as the film becomes thinner, the magnetization between the two rounded accumulations of topological charge progressively tilts toward the plane, $\theta_2 \to 90^\circ$. At the same time, the charge-carrying regions are pushed toward the central plane due to the reduced confinement, leading to an increase of $\theta_1$. 

The dependences $\theta_1(\nu)$ and $\theta_2(\nu)$ are shown in Fig.~\ref{fig08}(d).
These angles are evaluated along the solid blue curve in the phase diagram [Fig.~\ref{fig08}(e)], which has the same meaning as in Fig.~\ref{fig05}(b), but for fixed $k_u^s=40$: it denotes the zero--eigenvalue line of isolated CF--1 and, equivalently, the expansion boundary of the CF--1 phase.
The dashed blue curve in Fig.~\ref{fig08}(e) indicates the collapse of isolated CF--1 fingers.

At the critical reduced thickness $\nu \approx 0.8$, CF--1 collapses into the homogeneous state.
This critical thickness marks the instability of both isolated CF--1 solitons and their modulated phases and appears as a tricritical point in Fig.~\ref{fig08}(e), where the conical and homogeneous phases also meet.
The maximal topological charge density [inset in Fig.~\ref{fig08}(d)] decreases accordingly and vanishes at this point.
Importantly, this transition is not continuously connected to the homogeneous state: a finite topological structure is required to sustain the internal finger texture.

For the CF--2 phase, a comparable critical thickness terminates its stability [critical point in Fig.~\ref{fig08}(f)].
The modulated CF--2 phase exists to the left of the solid red curve in the phase diagram in Fig.~\ref{fig08}(f).
For small values of the uniaxial anisotropy, it is replaced by the conical phase [grey region in Fig.~\ref{fig08}(f)].
Notably, isolated bimerons remain metastable beyond the instability boundaries of the periodic CF--2 lattice, as indicated by the red shaded region bounded by the dashed red line in Fig.~\ref{fig08}(f).

\subsubsection{Effect of increasing film thickness}

The opposite tendency is observed with increasing film thickness.
According to Fig.~\ref{fig08}(g), the critical lines merge above $\nu \approx 1.2$.
As a result, the trivial CF--1 and unitary CF--2 fingers acquire very similar internal structures [Fig. \ref{fig08}(b),(c)].

To illustrate this, we conceptually divide the fingers by the central plane of the film.
The resulting upper (I and II) and lower (III and IV) parts can be interpreted as composite objects composed of coupled half-merons. Then, as shown in Fig.~\ref{fig08}(b), the upper halves of CF--1 and CF--2 become structurally indistinguishable.
Consequently, each upper part can combine with either lower part, yielding four distinct finger textures, as illustrated in Fig.~\ref{fig08}(c): I--III (CF-1), I--IV (CF-2), II--III (CF-2), and II--IV (CF-1).

Importantly, the type of the complete finger cannot be uniquely inferred from the upper half alone: the same upper configuration may correspond to different global textures depending on how it is coupled to the lower half.
In this sense, both CF--1 and CF--2 can be viewed as bimerons.

\begin{figure}
  \centering
  \includegraphics[width=0.99\linewidth]{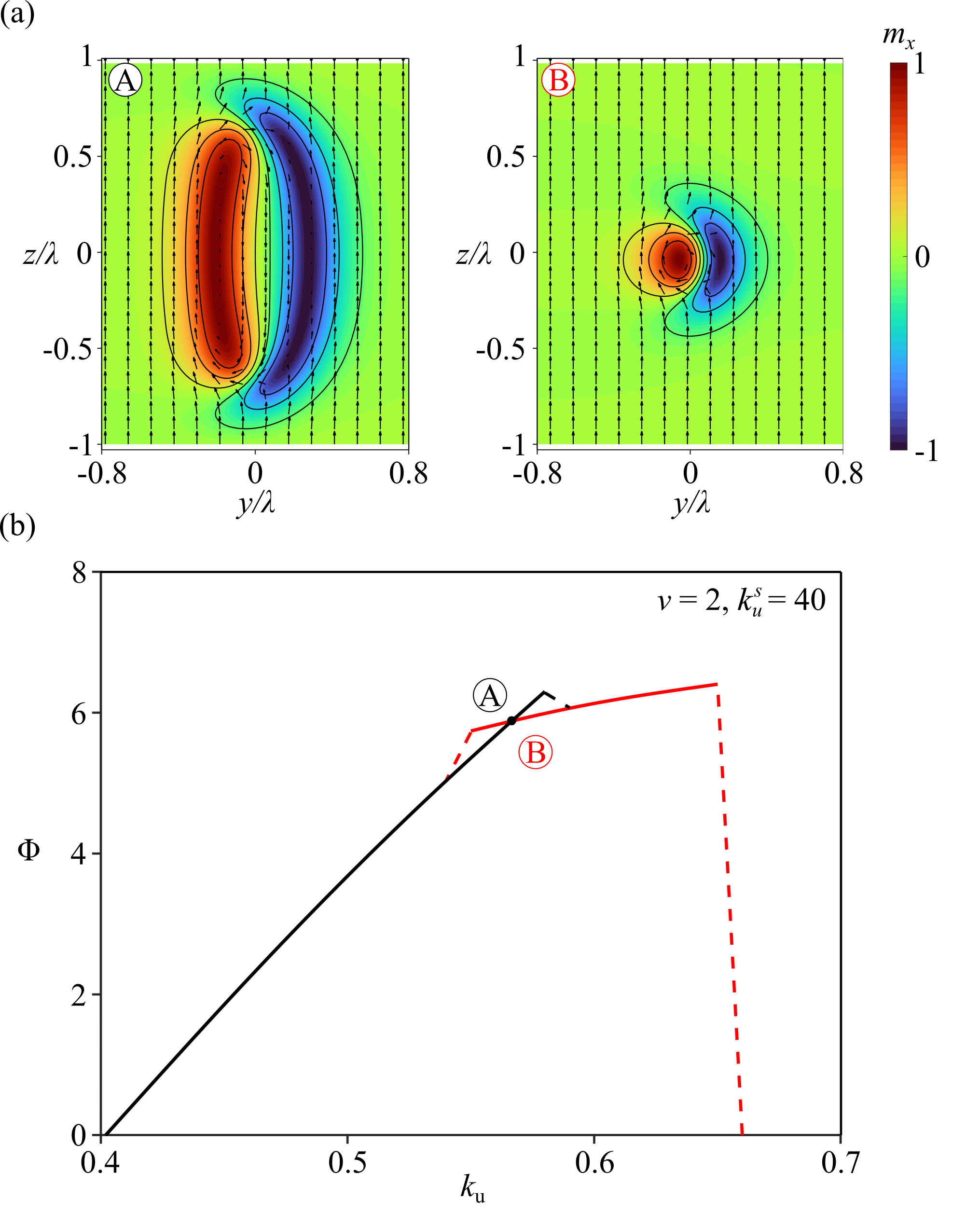}
\caption{\label{fig09}
Bistability of isolated bimerons at large film thickness.
(a) Color plots of the $m_x$ component of the magnetization for two bimeron configurations (large and small CF--2) obtained near the collapse point.
The smaller bimeron represents a bulk-like soliton weakly affected by surface pinning, whereas the larger one is stabilized by surface-induced anisotropy.
(b) Eigen-energy of isolated bimerons relative to the homogeneous state as a function of the uniaxial anisotropy.
The hysteresis loop indicates bistability, where large and small bimerons coexist within a finite range of anisotropy.
}
\end{figure}

\subsection{Bistability of isolated bimerons}
The anisotropy-driven evolution of isolated bimerons reveals yet another interesting feature at large film thicknesses [Fig.~\ref{fig09}].
In the immediate vicinity of the collapse point, a large isolated bimeron transforms into a smaller one.
Corresponding color plots of the $m_x$ component of the magnetization for the two bimeron varieties (A and B) are shown in Fig.~\ref{fig09}(a).
The small bimeron essentially represents a bulk-like soliton that would occur in chiral magnets in the absence of surface pinning.

The eigen-energy of bimerons, measured with respect to the homogeneous state, is plotted in Fig.~\ref{fig09}(b).
It exhibits a hysteresis loop, indicating bistability within a finite range of the uniaxial anisotropy.
In other words, the hysteretic behavior shown in Fig.~\ref{fig09}(b) reflects the coexistence of two distinct bimeron configurations with different sizes under the same external parameters.
Such bistability implies the presence of two local minima in the energy landscape and arises from the competition between surface-induced anisotropy and the bulk chiral interactions that stabilize the soliton core.
Physically, the large bimeron is strongly influenced by surface pinning, whereas the smaller configuration resembles a bulk-like soliton only weakly affected by the confining boundaries.

This bistability suggests potential applications of isolated bimerons as multistate solitonic information carriers.
In particular, the two stable configurations could encode different logical states under identical control parameters, enabling memory elements based on size-selective switching between large and small bimerons.
In addition, the difference in spatial extent between the two states may lead to distinct dynamical responses to spin torques, external fields, or defects, opening the possibility of racetrack-type architectures where information is encoded not only in the position of solitons but also in their internal state.
More generally, the coexistence of multiple metastable soliton branches highlights the rich energy landscape of confined chiral magnets and points toward new strategies for engineering functional magnetic textures through confinement and surface anisotropy.

\begin{figure*}
  \centering
  \includegraphics[width=0.99\linewidth]{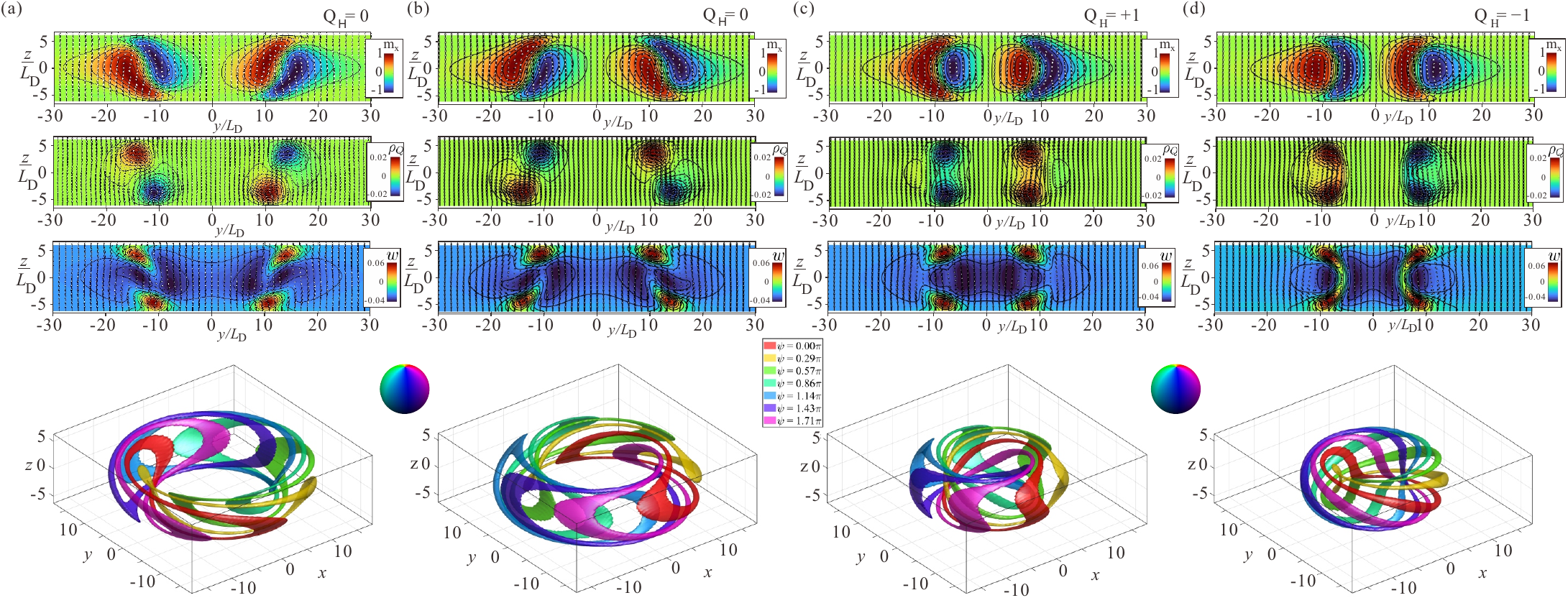}
  \caption{\label{fig10}
Four types of isolated hopfions obtained by revolving CF--1 and CF--2 textures about the $z$ axis. 
(a,b) Topologically trivial hopfions with Hopf index $Q_H=0$. 
(c,d) Hopfions with $Q_H=+1$ and $Q_H=-1$, respectively. 
Top panels show color maps of the $m_x$ component in $yz$ cross sections. 
Middle panels display the topological charge density and the total energy density [compare with the corresponding distributions for fingers in Fig. \ref{fig02}]. 
Bottom panels show preimages of the magnetization field plotted for the polar angle $\theta=\pi/2$ and several values of the azimuthal angle $\psi$. $k_u=0.22,\,k_u^s=50$.
}
\end{figure*}

\section{Isolated hopfions: precursor states of cholesteric fingers}

Isolated hopfions can be viewed as toroidal three-dimensional solitons obtained by revolving a two-dimensional CF--1 or CF--2 texture about the $z$ axis [Fig. \ref{fig10}]. 
Because four varieties of cholesteric fingers exist, this construction gives rise to four corresponding types of isolated hopfions.

From a topological perspective, hopfionic configurations are mappings
\(
\mathbf{m} : \mathbb{R}^3 \cup \{\infty\} \simeq S^3 \rightarrow S^2,
\)
which are classified by the third homotopy group~\cite{Faddeev,Bott}
\(
\pi_3(S^2) \cong \mathbb{Z}.
\)
Each homotopy class is characterized by an integer-valued Hopf invariant $Q_{\mathrm H}$.
Geometrically, this invariant represents the linking number of the preimages of two generic points on the order-parameter sphere, or equivalently the total linking of the field lines associated with the configuration.
It can be expressed as
\begin{equation}
    Q_{\mathrm H} = \frac{1}{(4\pi)^2} \int_{\mathbb{R}^3} 
    \mathbf{A}\cdot\mathbf{B}\, \mathrm{d}^3 x,
    \label{Hopf}
\end{equation}
where the emergent field
\(
\mathbf{B}
\)
is defined as $\mathbf{B}_i =
(1/2)
\epsilon_{ijk}\,
\mathbf{m}\cdot
\left(
\partial_j \mathbf{m}
\times
\partial_k \mathbf{m}
\right),$
and satisfies
\(
\nabla\!\cdot\!\mathbf{B}=0.
\)
Consequently, a vector potential $\mathbf{A}$ can be introduced such that $\nabla \times \mathbf{A} = \mathbf{B}$.

Hopfions generated from CF--1 textures have vanishing Hopf invariant, $Q_{\mathrm H}=0$ [Fig. \ref{fig10}(a),(b)].
Nevertheless, we retain the term \emph{hopfion} for these objects.
Following the terminology introduced in Ref.~\cite{ackerman2017}, the term is often used in a broader sense for three-dimensional toroidal solitons exhibiting closed-loop preimage structures.
The CF--1–derived configurations preserve the characteristic geometrical features associated with hopfions—namely a twisted toroidal core and mutually linked isosurfaces—even though their total Hopf charge is zero.
In this sense they represent geometrically hopfion-like states without net topological linking.

In contrast, hopfions constructed from CF--2 textures carry a nonzero Hopf charge [Fig. \ref{fig10}(c),(d)].
Depending on the sense of the bimeron rotation during the three-dimensional construction, the resulting Hopf invariant can be either positive or negative.

The phase diagram in Fig.~\ref{fig05}(e) provides a direct guideline for locating the stability window of metastable hopfions.
The eigen-energy of an isolated hopfion exhibits a well-defined minimum at an equilibrium radius $R_0$ that lies close to the phase boundaries of the corresponding finger states [see Ref. \cite{hopfions} for details].
This proximity indicates that hopfions represent localized precursor configurations of the modulated finger phases: they inherit the internal structure of cholesteric fingers but remain spatially confined.
Different hopfion varieties possess different equilibrium radii, which offers a practical experimental signature allowing them to be distinguished through their characteristic size.

The stability of these objects is controlled by the uniaxial anisotropy $k_u$.
As $k_u$ increases, all hopfionic configurations shrink and eventually collapse into toron-like states.
During this process their internal structure becomes progressively simplified and the distinction between the original hopfion varieties disappears, with all configurations converging toward a common toron topology.
In contrast, when the system approaches the stability regions of the CF--1 and CF--2 phases, the equilibrium radius grows rapidly and the hopfion expands without bound, signaling the transition toward extended finger textures.

When the magnitude of $k_u$ is reduced beyond the metastability window (i.e., inside the red- and blue-shaded regions in Figs.~\ref{fig05}(a),(b)), isolated hopfions become elliptically unstable. 
They elongate along one direction and gradually transform into extended textures that reconstruct the corresponding modulated finger phase whose equilibrium period was determined in Fig.~\ref{fig05}(f).
Under these conditions periodic lattices of hopfions cannot be stabilized. 
As shown in Ref.~\cite{hopfions}, the energy landscape of such lattices does not exhibit a local minimum and instead relaxes continuously toward the modulated finger phase.
The origin of this instability is essentially geometric. 
Once the energy of an individual finger becomes negative, the modulated finger phase can fill space with unit efficiency. 
In contrast, a lattice of hopfions necessarily contains extended regions of homogeneous background both inside the toroidal cores and in the interstitial space between neighboring objects. 
This inefficient packing increases the total energy and drives a deformation of the hopfionic textures: their vacuum cores collapse and the structures elongate, eventually evolving into finger-like configurations that tile the system more efficiently.

Although the present analysis clarifies the basic topology and energetics of isolated hopfions for thin films with thickness equal to one spiral pitch ($\nu/\lambda=1$), several aspects of their behavior in confined chiral systems remain open.
In particular, the evolution of hopfionic structures with increasing film thickness has not yet been systematically explored. 
As the confinement is relaxed, the internal three-dimensional structure of the toroidal core may undergo qualitative changes, potentially leading to transitions between surface-pinned and bulk-like configurations. 
Such effects could give rise to bistability between distinct hopfion states with different radii or internal structures, analogous to the bistable behavior observed for bimerons in thick films. 
Understanding these thickness-driven transformations and the associated energy landscape remains an important direction for future studies of hopfionic solitons in confined chiral media.

\section{Conclusions}

In this work we developed a unified theoretical description of cholesteric fingers in confined chiral liquid crystals within a continuum framework closely related to the Dzyaloshinskii theory of chiral magnets.
This correspondence allows the physics of cholesteric fingers to be interpreted in terms of topological solitons familiar from magnetic systems and establishes a direct conceptual bridge between chiral liquid crystals and chiral magnets.

Our analysis shows that the two principal finger varieties, CF--1 and CF--2, can be understood as composite chiral solitons formed by meronic constituents.
In particular, CF--2 corresponds to a bimeron with unit topological charge, while CF--1 represents a topologically trivial composite structure composed of two merons with identical vorticities.
Strong homeotropic surface anchoring plays a decisive role in stabilizing these textures and strongly modifies their internal structure by redistributing the topological charge across the film thickness.

Despite their composite character, isolated fingers embedded in the homogeneous background behave as particle-like objects interacting through repulsive forces arising from exponentially decaying asymptotics of the director field.
Periodic finger phases emerge when the eigen-energy of an isolated finger becomes negative relative to the homogeneous state, leading to nucleation-type phase transitions characterized by a diverging lattice period.
The presence of several energetically degenerate finger varieties enables the formation of mixed periodic sequences whose number grows combinatorially and can be classified using necklace enumeration and Jagodzinski-type local rules, closely analogous to stacking polytypes in close-packed crystals.

We further demonstrated that the character of the interaction between fingers depends strongly on the surrounding background state.
While fingers embedded in the homogeneous phase repel each other, the interaction becomes attractive in the conical background due to the overlap of distortion regions, allowing the formation of equilibrium bound configurations.

Finally, we analyzed the role of geometric confinement by varying the film thickness.
Reducing the thickness ultimately leads to the collapse of both CF--1 and CF--2 solitons at a critical value where their internal topological structure disappears.
In the opposite limit of large thickness, isolated bimerons exhibit bistability between surface-stabilized and bulk-like configurations, giving rise to a hysteretic energy landscape.

Taken together, these results demonstrate that cholesteric fingers constitute experimentally accessible realizations of composite chiral solitons whose behavior closely parallels that of magnetic textures while revealing additional phenomena induced by surface anchoring and confinement.
This unified perspective highlights chiral liquid crystals as a versatile platform for exploring soliton physics and suggests promising routes for engineering analogous structures in chiral magnetic systems through controlled surface anisotropies.

\bibliographystyle{apsrev4-2}
\bibliography{refs}

\end{document}